\documentclass[12pt]{article}
\usepackage{bbm}
\usepackage{graphics}
\usepackage{color}
\usepackage{authblk}
\usepackage{latexsym}
\usepackage{epsfig,amssymb,euscript, mathrsfs,cite}
\usepackage{amsmath}
\usepackage{slashed}
\usepackage{soul}
\usepackage{cancel}
\usepackage{verbatim}
\usepackage{mathrsfs} 
\usepackage[normalem]{ulem}
\usepackage{enumerate}
\definecolor{MyDarkBlue}{rgb}{0.15,0.15,0.45}
\usepackage[linktocpage=true]{hyperref}
\usepackage{bm}
\usepackage[new]{old-arrows}

\newcommand{\ii}{\mathrm{i}}
\newcommand{\ex}{\mathrm{e}}
\newcommand{\dd}{\mathrm{d}}

\newcommand{\R}{\mathbb{R}}
\newcommand{\C}{\mathbb{C}}
\newcommand{\Z}{\mathbb{Z}}

\newcommand{\cA}{\mathcal{A}}
\newcommand{\cB}{\mathcal{B}}
\newcommand{\cC}{\mathcal{C}}
\newcommand{\cD}{\mathcal{D}}
\newcommand{\cF}{\mathcal{F}}
\newcommand{\cK}{\mathcal{K}}
\newcommand{\cN}{\mathcal{N}}
\newcommand{\cU}{\mathcal{U}}
\newcommand{\cX}{\mathcal{X}}

\newcommand{\AABJM}{\cA_{\rm ABJM}}
\newcommand{\Cthree}{\mathscr{C}}
\newcommand{\LM}{\mathcal{L}_{\rm M}}
\newcommand{\rflux}{\varrho}

\newcommand{\mc}[1]{\mathcal{#1}}
\newcommand{\mf}[1]{\mathfrak{#1}}

\newcommand{\zetanew}{\xi}
\newcommand{\Phianom}{\Phi^{(\text{anom})}}
\newcommand{\cnew}{\mathfrak{c}}

\newcommand{\hook}
{\mathbin{\rule[.2ex]{.4em}{.03em}\rule[.2ex]{.03em}{.9ex}}}
\newcommand{\nn}{\notag}

\hypersetup{
colorlinks=true,
citecolor=MyDarkBlue,
linkcolor=MyDarkBlue,
urlcolor=MyDarkBlue,
pdfauthor={},
pdftitle={},
pdfsubject={hep-th}
}
\newsavebox{\ns}
\newsavebox{\dbrane}
\newsavebox{\dbshort}

\newcommand\diff{\mathrm{d}}

\newcommand{\rd}{{\rm{d}}}

\newlength{\sswidth}

\numberwithin{equation}{section}       

\begin{document}

\pagestyle{plain}
\setcounter{page}{1}
\newcounter{bean}
\baselineskip18pt

\begin{titlepage}

\vfill



\begin{center}
   \baselineskip=16pt
   {\Large\bf 
Localizing the M-Theory Path Integral}
  \vskip 1cm
Pietro Benetti Genolini,$^1$ Florian Gaar,$^2$  Jerome P. Gauntlett,$^3$ \\Jaeha Park$^3$ 
and James Sparks$^{2}$ \\
     \vskip 1cm     
                                                    \begin{small}
                                \textit{$^1$D\'epartement de Physique Th\'eorique,
Universit\'e de Gen\`eve,\\
24 quai Ernest-Ansermet,
1211 Gen\`eve, Suisse}
        \end{small}\\
                        \begin{small}\vskip .3cm
      \textit{$^2$    Mathematical Institute,
University of Oxford,\\
Woodstock Road,
Oxford OX2 6GG, U.K.}
        \end{small}\\
        \begin{small}\vskip .3cm
      \textit{$^3$Abdus Salam Centre for Theoretical Physics, 
  Imperial College,\\ Prince Consort Rd., London, SW7 2AZ, U.K.}
        \end{small}\\
                       \end{center}
\vfill

\begin{center}
\textbf{Abstract}
\end{center}

\begin{quote}
We propose a supersymmetric localization formula for the local perturbative
contribution to the M-theory path integral, reducing it to an integral over
equivariant flux data. Relative equivariant extensions of the eleven-dimensional Chern--Simons
couplings determine the localized action, while for M2-brane backgrounds a tangent-cotangent
character computes the one-loop determinant. This
gives the first closed form expression for the complete perturbative ABJM partition function at general
squashing and chemical potentials, and new analytic results for the ABJM topologically twisted index describing black holes. Applied to the
non-toric $V^{5,2}$ theory, the same prescription yields a closed analytic
Airy constant previously known only numerically.
Our formulas reproduce all available independent field theory results, and provide analytic predictions at general parameters where no field theory answer is currently known.
We derive an all-genus result for the Type IIA expansion 
which is an equivariant form of the usual topological string constant map result, and also extract integrated correlators. 
\end{quote}

\vfill

\end{titlepage}

\tableofcontents

\newpage

\section{Introduction}

Supersymmetric localization provides a powerful way of computing quantum field theory path integrals exactly \cite{Witten:1988ze,Nekrasov:2002qd,Pestun:2007rz,Pestun:2016zxk}. 
The path integral reduces to an integral over a supersymmetric locus, together with a one-loop determinant governed by a corresponding cohomological complex. 
For strongly coupled supersymmetric field theories this has led to exact results that are inaccessible by conventional perturbative methods. 
A natural question is whether an analogous principle can be applied directly to quantum gravity, and in particular to M-theory.

We consider this question in a holographic setting, where the asymptotic boundary conditions provide a natural way to select the supersymmetry with respect to which the path integral should localize \cite{deWit:2018dix}. We take the physical eleven-dimensional spacetime $M$ to have a conformal boundary $\partial_\infty M$, and fix asymptotic data preserving a supercharge $\mathcal{Q}$. 
 A bulk Killing spinor approaches the corresponding prescribed boundary Killing spinor, while its bilinear defines a supersymmetric Killing vector $K$. The vector $K$ generates the equivariant action \cite{BenettiGenolini:2023kxp} that plays a key role throughout our construction. 
 We propose that the local perturbative contribution to the M-theory path integral takes the form
 \begin{align}
 \label{eq:master_2}
	 Z_{\rm M}^{\rm pert}[\mathscr B] &=\int [\dd\nu] \, \exp\!\left[-I_{\rm SUSY}\right]  Z_{\text{1-loop}} \, .
\end{align}
Equation \eqref{eq:master_2} has the general form of a supersymmetric
localized partition function, with the exponential of the supersymmetric
action multiplied by a one-loop factor and integrated over the localized
BPS data. In the holographic setting the asymptotic data $\mathscr B$ are
held fixed, leaving only bulk equivariant data not determined at the
conformal boundary to be integrated over. We denote these relative data
collectively by $\nu$. Thus the $\nu$ integral is the remnant of the bulk
supergravity path integral after localization; its precise meaning will
become clear below once the relevant equivariant flux variables are
introduced.\footnote{In general there is also a sum over localized
saddles in \eqref{eq:master_2}, as well as non-perturbative contributions
that we do not consider in this work.}

The exponent $I_{\rm SUSY}$ in \eqref{eq:master_2} is the supersymmetric contribution of the
supergravity action, evaluated using relative equivariant localization 
\cite{BenettiGenolini:2023kxp,BenettiGenolini:2026cdw,BenettiGenolini:2026cyc}. For the M-theory Chern--Simons sector
this requires an auxiliary twelve-dimensional filling $W$ of $M$, together
with an extension of the $C$-field curvature over $W$
\cite{Witten:1996md}. Subject to the conjectures of
\cite{BenettiGenolini:2026cdw,BenettiGenolini:2026cyc}, this construction captures perturbative
corrections to all orders in the Planck length, with non-trivial holographic
evidence provided in \cite{BenettiGenolini:2026cyc}.

The new ingredient is our proposal for the one-loop factor
$Z_{\text{1-loop}}$. As in ordinary supersymmetric localization, we expect it
to be determined by the equivariant character of the quadratic
$\mathcal Q$-complex governing fluctuations about the localized saddles.
We propose that the relevant equivariant character is naturally formulated on
the same twelve-dimensional filling $W$ that enters the relative definition of
the M-theory action. 
We make this proposal concrete for a class of holographic M2-brane backgrounds, with eleven-dimensional geometries that locally asymptote to 
$AdS_4\times Y_7$. Here $Y_7$ is a Sasaki--Einstein seven-manifold and hence the link of a Calabi--Yau four-fold cone. The twelve-dimensional filling $W$ then naturally splits into a $4+8$ decomposition. In this setting we take the relevant character to be the 
localized equivariant Dolbeault character associated with the virtual bundle
$TW-T^*W$ (described in more detail below \eqref{eq:character-general}). 
This form is motivated by twisted formulations of eleven-dimensional
supergravity \cite{Costello:2016mgj,Costello:2018zrm,Raghavendran:2021qbh,Hahner:2023kts} and protected M-theory indices
\cite{Nekrasov:2014nea}.

We use this construction to compute squashed three-sphere partition functions
of holographic M2-brane theories, including ABJM and the theory dual to
$AdS_4\times V^{5,2}$. Supersymmetric localization in the dual field theories
leads to a perturbative grand potential of the cubic form
\begin{align}\label{Jpert}
J^{\rm pert}(\mu)
&=
\frac{\cC}{3}\mu^3+\cB\mu+\cA\,,
\end{align}
where $\mu$ is the grand canonical chemical potential conjugate to the
M2-brane number $N$. Equivalently, the canonical partition function takes
the familiar Airy form
\begin{align}
\label{eq:Zpert_Airy}
Z^{\rm pert}(N)
&=
\cC^{-1/3}\mathrm{e}^{\cA}
\operatorname{Ai}\!\left[\cC^{-1/3}(N-\cB)\right]\,,
\end{align}
up to non-perturbative corrections
\cite{Fuji:2011km,Marino:2011eh,
Bobev:2023lkx,Bobev:2025ltz}.
This Airy form resums the full perturbative large $N$
expansion around the leading $N^{3/2}$ behaviour.

In \cite{BenettiGenolini:2026cyc}, the localized action
$I_{\rm SUSY}$ reproduced the coefficients $\cC$ and $\cB$ directly from
eleven-dimensional M-theory, while the constant $\cA$ remained undetermined.
Here $Z_{\text{1-loop}}$ supplies precisely this missing
piece. Our construction leads to analytic expressions for $\cA$, as a function of general 
squashing and flavour chemical potentials, and agrees with all available 
field theory results in the special cases where these are known.
For ABJM theory, the type IIA large $k$ expansion 
reproduces, directly from eleven dimensions, the universal Bernoulli/Hodge coefficients of higher-genus topological string constant maps.

We also apply the prescription to the topologically twisted index of ABJM, extending the analysis of \cite{BenettiGenolini:2026cyc} to
include $Z_{\text{1-loop}}$. The same local one-loop block for the squashed three-sphere partition function appears at the two
fixed points of the corresponding Euclidean black hole geometries.
We determine closed form expressions for the $N$-independent constant terms that appear in the all-orders $1/N$ expansion of the topologically twisted index,
for generic flavour chemical potentials and magnetic charges.

\section{The localized M-theory path integral}\label{locpisec}
We begin by recalling the computation of the semi-classical part of the M-theory path integral as in
\cite{BenettiGenolini:2026cdw,BenettiGenolini:2026cyc}. The $D=11$ action consists of a gauge-invariant part, which is an infinite expansion in the Planck length $\ell_p$,
along 
with the anomaly-fixed Chern--Simons terms, 
which are known exactly. 
To define the latter one picks a twelve-manifold filling
$W$ with $M=\partial W$ as a boundary face, extending also the four-form $G = \dd C$ \cite{Witten:1996md}.
An essential ingredient is an equivariantly closed completion of $G$, $\Phi^{G}$, extended from $M$ to $W$ with
$\dd_K \Phi^G = 0$, where $\dd_K \equiv \dd - K \hook$ is the equivariant exterior derivative.
The equivariantly closed anomaly form $\Phianom$ on $W$ is 
\begin{equation}
	\Phianom = \frac{\ii}{6} (\Phi^G)^3 + \ii (2\pi\ell_p)^6 \Phi^G \wedge \Phi^{X_8} \,,
\end{equation}
where the second term comes from the well-known eight-derivative correction \cite{Vafa:1995fj,Duff:1995wd}.
The equivariant completion $\Phi^{X_8}$ of the eight-form $X_8 = \frac{1}{192} (P_1^2 - 4 P_2)$ is
defined by replacing the Riemann curvature two-form $R^{ab}$ by $R^{ab} - \frac{1}{2} (\dd K^\flat)^{ab}$
in the Pontryagin forms $P_1, P_2$.

In \cite{BenettiGenolini:2026cdw,BenettiGenolini:2026cyc}, it was argued that the supersymmetric action $I_{\text{SUSY}}$ can be evaluated 
using the Berline--Vergne--Atiyah--Bott (BVAB) fixed point theorem \cite{BV:1982, Atiyah:1984px}. Assuming that the fixed points of $K$ are in the interior of $W$ we have
\begin{align}\label{ISUSYloc}
I_{\text{SUSY}}  =\frac{1}{(2\pi)^8\ell_p^9} \int_{\mathscr{F}} \frac{\Phianom}{e_{K}(N\mathscr{F})}\, .
\end{align}
Here $\mathscr{F} \equiv\{K=0\}\subset W$ is the fixed point set of 
$K$ in $W$, with normal bundle $N\mathscr{F}$ and associated equivariant Euler class $e_{K}$.
 While we have only proved this at two-derivative level, we have conjectured that \eqref{ISUSYloc} is an exact result valid to all orders in $\ell_p$, based on holographic evidence \cite{BenettiGenolini:2026cyc,BenettiGenolini:2026cdw}. 

We now turn to the new ingredient $Z_{\text{1-loop}}$. The
tangent-cotangent structure is motivated by twisted eleven-dimensional
supergravity \cite{Raghavendran:2021qbh}. For the minimal twist on a Calabi--Yau five-fold $\cX$ times a
real line, the free $\mathcal Q$-complex has local supercharacter given,
up to the overall grading convention, by the Dolbeault character of
$T\cX-T^*\cX$ \cite{Raghavendran:2021qbh}. We propose that for the holographic
M2-brane backgrounds considered here, after imposing the supersymmetric
and relative boundary conditions, the corresponding unpaired
fluctuations are encoded by a relative twelve-dimensional
tangent-cotangent character on the same filling $W$ that enters the
Chern--Simons construction. Thus, in parallel with
\eqref{ISUSYloc}, for the M2-brane backgrounds of this
paper we define 
\begin{align}
    \mathcal K_W(t)
    &\equiv
    \operatorname{Ind}^{\rm rel}_{2\pi tK}
    \!\left[\bar\partial_W\otimes(TW-T^*W)\right]
    \nonumber\\
    &=
    \int_{\mathscr F}
    \frac{
    \operatorname{ch}_{2\pi tK}(TW-T^*W)\,
    \operatorname{Td}_{2\pi tK}(TW)}
    {e_{2\pi tK}(N\mathscr F)} \, .
    \label{eq:character-general}
\end{align}
Here $TW$ denotes the holomorphic tangent bundle $T^{1,0}W$, and
$T^*W$ its holomorphic cotangent bundle. More explicitly, the
tangent-cotangent character is the equivariant index of the virtual Dolbeault complex
$\Omega^{0,\bullet}(W,TW)-\Omega^{0,\bullet}(W,T^*W)$, with the
appropriate relative boundary conditions. 
We assume that the
equivariant action preserves the chosen complex structure and use the
induced complex orientation.

For compact $W$, the second line of \eqref{eq:character-general} is the equivariant Hirzebruch--Riemann--Roch formula
followed by BVAB localization. In the non-compact setting relevant here, we
instead regard the fixed locus expression as defining the corresponding
localized relative character. The one-loop factor is then obtained by zeta
regularization, 
with the spectral zeta function defined by the Mellin transform
\begin{align}
    \zeta_{\mathcal K}(s)
    &=
    \frac{1}{\Gamma(s)}
    \int_0^\infty \dd t\,t^{s-1}\mathcal K_W(t)\, ,
    \nonumber\\
    \log Z_{\text{1-loop}}
    &=
    \zeta'_{\mathcal K}(0)\, ,
    \label{eq:zeta}
\end{align}
where meromorphic continuation in $s$ is
understood.
We remark that in writing \eqref{ISUSYloc}, \eqref{eq:character-general}, and the formulas in the remainder of this section,
we have assumed that $W$ is a smooth manifold; the ABJM orbifold example will be treated using a Molien projection. 
The specific holomorphic form of our one-loop character \eqref{eq:character-general} is appropriate to the class of complex backgrounds we consider; more general supersymmetric saddles need not admit such a global complex description, a point we return to in section \ref{sec:Discussion}. 

For the backgrounds considered below we assume that the fixed points of $K$
in $W$ are isolated, and label them by $A\in\mathscr F$. Let
$w_{A,\alpha}$, $\alpha=1,\ldots,6$, denote the additive weights of $K$ on
$T_AW$. At an isolated fixed point \cite{BenettiGenolini:2026cdw,BenettiGenolini:2026cyc}
\begin{align}
 e_K(T_AW)
 &=
 \frac{1}{(2\pi)^6}
 \prod_{\alpha=1}^{6}w_{A,\alpha}\, ,
\nn\\
 (\Phi^{X_8})_0|_A
 &=
 \frac{1}{192(2\pi)^4}
 \Bigg[
 \left(\sum_{\alpha=1}^{6}w_{A,\alpha}^2\right)^2
 -4\sum_{\alpha<\beta}
 w_{A,\alpha}^2w_{A,\beta}^2
 \Bigg]\, .
 \label{eq:x8fixed}
\end{align}
Thus the localized action \eqref{ISUSYloc} becomes
\begin{align}
 &I_{\rm SUSY}
  =
 \sum_A
 \frac{\ii(\Phi^G)_0|_A}
 {\prod_{\alpha=1}^{6}w_{A,\alpha}}
 \Bigg\{
 \frac{\big[(\Phi^G)_0|_A\big]^2}
 {24\pi^2\ell_p^9}
 +
 \frac{1}{192\ell_p^3}
 \Big(\sum_{\alpha=1}^{6}w_{A,\alpha}^2\Big)^2-
 \frac{1}{48\ell_p^3}
 \sum_{\alpha<\beta}
 w_{A,\alpha}^2w_{A,\beta}^2
 \Bigg\}\, ,
 \label{eq:ISUSYfixed}
\end{align}
where $(\Phi^G)_0|_A$ denotes the zero-form part of $\Phi^G$ 
at the fixed point $A\in\mathscr{F}$.

For the one-loop character we introduce the corresponding multiplicative weights
\begin{equation}
 q_{A,\alpha}\equiv \mathrm{e}^{-t w_{A,\alpha}}\, ,
 \label{eq:qweights}
\end{equation}
with $t$ the equivariant parameter.
The condition that the Killing spinor is uncharged under $K$ implies
\begin{equation}
 \sum_{\alpha=1}^{6}w_{A,\alpha}=0
 \quad\Longleftrightarrow\quad
 \prod_{\alpha=1}^{6}q_{A,\alpha}=1\, .
 \label{eq:CYweights}
\end{equation}
At an isolated fixed point $A$, let
\begin{align}
 x_{A,\alpha}
 &=
 t\mskip1mu w_{A,\alpha}\,,
 \qquad
 q_{A,\alpha}
 =
 \ex^{-x_{A,\alpha}}\,.
\end{align}
The equivariant Euler class, Todd class, and tangent-cotangent Chern
character appearing in \eqref{eq:character-general} are then
\begin{align}
 e_{2\pi tK}(T_AW)
 &=
 \prod_{\alpha=1}^{6}x_{A,\alpha}\,,\qquad
 \operatorname{Td}_{2\pi tK}(T_AW)
 =
 \prod_{\alpha=1}^{6}
 \frac{x_{A,\alpha}}{1-\ex^{-x_{A,\alpha}}}\,,
 \nonumber\\
 \operatorname{ch}_{2\pi  tK}(T_AW-T_A^*W)
 &=
 \sum_{\alpha=1}^{6}
 \left(
 \ex^{x_{A,\alpha}}-\ex^{-x_{A,\alpha}}
 \right)\,.
\end{align}
The general expression \eqref{eq:character-general} then reduces to
\begin{equation}
 \mathcal K_W(t)
 =
 \sum_{A\in\mathscr F}
 \frac{\sum_{\alpha=1}^{6}
 (q_{A,\alpha}^{-1}-q_{A,\alpha})}
 {\prod_{\alpha=1}^{6}
 (1-q_{A,\alpha})}\, .
 \label{eq:character}
\end{equation}
The zero-charge condition also implies
$\mathcal K_W(-t)=-\mathcal K_W(t)$, and hence $\zeta_{\mathcal K}(0)=0$.\footnote{For the general Dolbeault character 
\eqref{eq:character-general} this is the
equivariant Serre duality relation in complex dimension six: the canonical
bundle is equivariantly trivial, while
$(TW-T^*W)^*=-(TW-T^*W)$.} 
A common rescaling $K\to\lambda K$ sends
$\mathcal K_W(t)\to\mathcal K_W(\lambda t)$ and therefore
$\zeta_{\mathcal K}(s)\to\lambda^{-s}\zeta_{\mathcal K}(s)$, so
$\log Z_{\text{1-loop}}=\zeta'_{\mathcal K}(0)$ is independent of the overall
normalization of $K$.

For the M2-brane backgrounds we study $W$ is the total space of a
fibration 
\begin{equation}\label{fibredform}
 Z_8 \longhookrightarrow W \longrightarrow M_4\, ,
\end{equation}
where $Z_8$ is a Calabi--Yau four-fold with $\partial Z_8=Y_7$, while $M_4$
is the non-compact four-dimensional spacetime with holographic boundary
$M_3=\partial_\infty M_4$. 
When the cone $C(Y_7)$ admits a smooth equivariant crepant resolution, the localized character $\mathcal K_W(t)$ is independent of the choice of resolution, by crepant birational invariance of the equivariant elliptic class and its Hirzebruch specialization \cite{BorisovLibgober2003}. We comment on this further in section \ref{sec:Discussion}.

With this $4+8$ decomposition we refine the fixed point label as
$A=(p,a)$, where $p\in M_4$ and $a\in Z_8$ label fixed points of the base
and fibre, respectively, and split the six tangent weights as
\begin{equation}
 \big(w_{p,a,1},\ldots,w_{p,a,6}\big)
 =
 \big(b_{p,1},b_{p,2},
 u_{p,a,1},\ldots,u_{p,a,4}\big)\, .
 \label{eq:weightsplit}
\end{equation}
The zero-charge condition \eqref{eq:CYweights} is therefore
\begin{equation}
 b_{p,1}+b_{p,2}
 +\sum_{I=1}^{4}u_{p,a,I}=0\, .
 \label{eq:zerochargesplit}
\end{equation}
Following \cite{BenettiGenolini:2026cyc}, it is then convenient to
introduce the local equivariant $G$-flux variables
\begin{equation}
 \mu_{p,a}
 \equiv
 \frac{\ii\,(\Phi^G)_0|_{p,a}}
 {b_{p,1}b_{p,2}\ell_p^3}\, .
 \label{eq:mudef}
\end{equation}
For each internal fixed point $a$, equivariant integration
of the $G$-flux over $M_4$ defines
\begin{align}
 \mu_a
 &\equiv
 \frac{\ii}{(2\pi)^2\ell_p^3}
 \left.\int_{M_4}^{\rm eq}\Phi^G\right|_a
 =
 \sum_{p\in M_4^K}\mu_{p,a}\, ,
 \label{eq:mutotal}
\end{align}
where $M_4^K$ denotes the fixed points of $K$ in $M_4$. 
Thus equivariant integration over $M_4$ maps the local flux
data $\{\mu_{p,a}\}$ to the integrated fluxes $\{\mu_a\}$,
labelled by the internal fixed points $a$.
We restrict attention to configurations for which these
integrated fluxes are independent of the internal fixed point
label $a$ and set
\begin{equation}
 \mu_a\equiv\mu\, ,\qquad \text{for all } a\, .
 \label{eq:diagonalmu}
\end{equation}
The common value $\mu$ is the M2-brane chemical potential and
is held fixed as part of the asymptotic boundary data
$\mathscr{B}$ in \eqref{eq:master_2}.
Allowed variations of the local flux data at fixed $\mu$
lie in the kernel of the equivariant integral
\eqref{eq:mutotal}, and hence satisfy
\begin{equation}
 \sum_{p\in M_4^K}\delta\mu_{p,a}=0
 \qquad \text{for all } a\, .
 \label{eq:mukernel}
\end{equation}
The independent continuous relative parameters remaining
after imposing the other boundary conditions and
flux quantization constraints are denoted collectively by
$\nu$ and are integrated over in \eqref{eq:master_2}.
More general configurations involving baryonic chemical potentials
and additional quantized $G$-flux data will not be considered
here.\footnote{In more general cases $\cA$ may effectively receive contributions from both the localized action and the one-loop determinant, and more delicate global phase data may also enter in~\eqref{eq:master_2}. }

The simplest case is Euclidean $AdS_4$, $M_4=EAdS_4\cong \C^2$, which has a single fixed point $p$ at the origin.
Since there is a single external fixed point, \eqref{eq:mutotal} and \eqref{eq:diagonalmu} imply $\mu_{p,a}=\mu$ for every internal fixed point $a$. There are therefore no remaining relative flux variables $\nu$ to integrate over.
The localized partition function \eqref{eq:master_2}
therefore reduces to
\begin{align}
\label{eq:Zpert_Jpert}
	Z^{\rm pert}_{\rm M}[\mu] & = \exp\big( - \left. I_{\rm SUSY}\right|_{\mu_a=\mu} \big)  Z_{\text{1-loop}} \equiv \exp \left( J_{\rm M}^{\rm pert}(\mu) \right) \, .
\end{align}
Here we are computing in the grand canonical ensemble for
the $C$-field
\cite{Gautason:2025plx,vanMuiden:2026nsp,Bobev:2026gir,
BenettiGenolini:2026cyc}.
Comparing with the results of supersymmetric localization,
we identify $J_{\rm M}^{\rm pert}(\mu)$ with the perturbative
grand potential \eqref{Jpert} associated with the holographic
dual field theory on $\partial_\infty M_4$, a
$U(1)\times U(1)$-squashed three-sphere.
Substituting \eqref{eq:mudef} into \eqref{eq:ISUSYfixed}
determines $\cC$ and $\cB$, as in
\cite{BenettiGenolini:2026cyc}, while
$\cA=\zeta'_{\mathcal K}(0)$.
An inverse Laplace transform in $\mu$, along a suitable
contour, then gives the canonical partition function
\eqref{eq:Zpert_Airy} at fixed $N$.
The squashing is determined by the ratio $b_1/b_2$;
we use the conventions
\begin{equation}
 b=\sqrt{\frac{b_1}{b_2}}\, ,\qquad Q=b+\frac{1}{b}\, .
\end{equation}

\section{ABJM theory}\label{secabjm}

We first apply the prescription of the previous section to the squashed three-sphere partition
function of the $U(N)_k \times U(N)_{-k}$ ABJM theory \cite{Aharony:2008ug}, dual to M-theory on $AdS_4 \times S^7/\Z_k$. For Chern--Simons level $k$ we take
\begin{equation}
 W= \C^2\times(\C^4/\Z_k)\, ,
\end{equation}
with a single fixed point at the origin. The six tangent weights are
$(b_1,b_2,u_1,\ldots,u_4)$. 
The bulk $U(1)^4$ equivariant weights are related to the boundary flavour chemical potentials by\footnote{This relation may be viewed as the eleven-dimensional uplift of the UV-IR relation derived in $D=4$ gauged supergravity in \cite{BenettiGenolini:2024lbj}.}
\begin{equation}
 \Delta_I=-\frac{2u_I}{b_1+b_2}\, .
 \label{eq:ABJMvariables}
\end{equation}
The zero-charge condition implies $\sum_{I=1}^4\Delta_I=2$.
The relation between these chemical potentials and real mass deformations
around the superconformal point is given in~\eqref{eq:ABJMmassmap}.

For the flux sector considered here the localized action has no
$\mu$-independent term. Substituting the ABJM fixed point data into
\eqref{eq:ISUSYfixed} gives~\cite{BenettiGenolini:2026cyc}
\begin{align}
 \cB&=
 \frac{1}{24k}
 \left[
 1-2\sum_I\frac{1}{\Delta_I}
 +\frac{4}{Q^2}\sum_{I<J}
 \frac{1}{\Delta_I\Delta_J}
 \right]
 +\frac{k^2-1}{24k}\, , \nonumber\\
 \cC&=
 \frac{2}
 {\pi^2 k Q^4\prod_{I=1}^4\Delta_I}\, .
 \label{eq:ABJMCB}
\end{align}
The first contribution to $\cB$ is generated by the $X_8$ term, while the
last term 
accounts for the anomalous M2-brane charge of the $\mathbb C^4/\mathbb Z_k$ orbifold singularity 
\cite{Bergman:2009zh,BenettiGenolini:2026cyc}.

We next turn to the one-loop character. For $k=1$ the orbifold is absent,
and \eqref{eq:character} gives
\begin{equation}
 \mathcal K_1(t)
 =
 \frac{\sum_{\alpha=1}^{6}
 (q_\alpha^{-1}-q_\alpha)}
 {\prod_{\alpha=1}^{6}(1-q_\alpha)}\, ,
 \qquad
 q_\alpha=\mathrm e^{-t w_\alpha}\, ,
 \label{eq:ABJMk1}
\end{equation}
where 
$ (w_1,\ldots,w_6)=(b_1,b_2,u_1,u_2,u_3,u_4)$. 
Using the rescaling freedom discussed below \eqref{eq:character}, we choose
\begin{equation}
 b_1=b\, ,\qquad b_2=b^{-1}\, ,
 \qquad
 u_I=-\frac{Q\Delta_I}{2}\, ,
 \label{eq:ABJMnormalizedweights}
\end{equation}
and we use this normalization for all $k$.

For general $k$, the $\Z_k$ action is non-trivial only on the four
internal complex directions, with charges
\begin{equation}
 (c_1,\ldots,c_6)=(0,0,1,1,-1,-1)\, ,
 \qquad
 \omega=\mathrm e^{2\pi\ii/k}\, .
\end{equation}
The one-loop character is obtained by projecting onto
$\Z_k$-invariant fluctuation modes, with corresponding Molien average \cite{Vergne1996EQUIVARIANTIF}: 
\begin{align}
 \mathcal K_k(t)
 &=
 \frac{1}{k}\sum_{r=0}^{k-1}
 \frac{ \sum_{\alpha=1}^{6}
 \left(
 \omega^{-r c_\alpha}q_\alpha^{-1}
 -\omega^{r c_\alpha}q_\alpha
 \right)}
 { \prod_{\alpha=1}^{6}
 \left(1-\omega^{r c_\alpha}q_\alpha\right)}\, .
 \label{eq:ABJMMolien}
\end{align}
The small $t$ expansion of this character begins as $\mathcal K_k(t)
=\mathcal K_{k,-3}t^{-3}
+\mathcal K_{k,-1}t^{-1}
+O(t)$,
so its Mellin representation initially converges for $\operatorname{Re}s>3$; meromorphic continuation to $s=0$ is understood below.\footnote{\label{footfive}One way to implement the meromorphic continuation is to split the Mellin integral at an arbitrary $T$, subtract the singular small $t$ terms on $(0,T)$, integrate these analytically, and then take $s\to0$. The resulting finite part is independent of $T$.}

To evaluate the Mellin transform we first work in a chamber with $\operatorname{Re}(Q\Delta_I)>0$, so that the geometric series may be written in terms of decaying exponentials. This gives Barnes periods $Q\Delta_I/2$. The resulting expression is then analytically continued to general chemical potentials.
 Define
\begin{equation}
 \bm\Omega_6=
 \left(
 b,b^{-1},
 \frac{Q\Delta_1}{2},\ldots,
 \frac{Q\Delta_4}{2}
 \right)\, .
 \label{eq:ABJMOmega}
\end{equation}
For $k=1$, writing
\begin{equation}
 D_+(t)=(1-\mathrm e^{-bt})(1-\mathrm e^{-b^{-1}t})
 \prod_{I=1}^4\left(1-\mathrm e^{-Q\Delta_I t/2}\right)\, ,
 \label{eq:ABJMDplus}
\end{equation}
the identity
$1-\mathrm e^{Q\Delta_I t/2}
=-\mathrm e^{Q\Delta_I t/2}
(1-\mathrm e^{-Q\Delta_I t/2})$, together with
$\sum_I\Delta_I=2$, gives
\begin{align}
 \mathcal K_1(t)
 &=\frac{1}{D_+(t)}\Bigg\{
 \sum_{\epsilon=b,b^{-1}}
 \left[\mathrm e^{-\epsilon t}
 -\mathrm e^{-(Q+\epsilon)t}\right]
 +\sum_{I=1}^4\left[
 \mathrm e^{-Q(1+\Delta_I/2)t}
 -\mathrm e^{-Q(1-\Delta_I/2)t}\right]
 \Bigg\}\, .
 \label{eq:ABJMk1decay}
\end{align}
Since 
$ {\mathrm e^{-zt}}/{D_+(t)}
 =\sum_{\bm m\in\Z_{\geq0}^{6}}
 \mathrm e^{-t(z+\bm m\cdot\bm\Omega_6)}$, 
the Mellin transform of \eqref{eq:ABJMk1decay} is immediately expressed
in terms of six-fold Barnes zeta functions. For general $k$, the same
geometric series expansion may be performed in each Molien sector, with
the roots of unity weighting the corresponding monomials. We hence
introduce the twisted Barnes zeta function
\begin{equation}
 \zeta_6^{(r)}(s,z)
 \equiv
 \sum_{\bm m\in\Z_{\geq0}^{6}}
 \frac{
 \omega^{-r\sum_{I=1}^{4}c_{I+2}m_{I+2}}
 }
 {(z+\bm m\cdot\bm\Omega_6)^s}\, .
 \label{eq:twistedBarnes}
\end{equation}
Term-by-term Mellin transformation of \eqref{eq:ABJMMolien} then gives
\begin{align} 
 \AABJM
=\frac{1}{k}\partial_s\sum_{r=0}^{k-1}\Bigl\{
&\sum_{\epsilon=b,b^{-1}}
\bigl[\zeta_6^{(r)}(s,\epsilon)
-\zeta_6^{(r)}(s,Q+\epsilon)\bigr]
+\sum_{I=1}^{4}\omega^{-r c_{I+2}}
\zeta_6^{(r)}\!\Big(
s,Q\Big(1+\frac{\Delta_I}{2}\Big)\Big)
\nonumber\\
 -&\sum_{I=1}^{4}\omega^{r c_{I+2}}
\zeta_6^{(r)}\!\Big(
s,Q\Big(1-\frac{\Delta_I}{2}\Big)\Big)
\Bigr\}\Big|_{s=0}\, ,
\label{eq:ABJMBarnes}
\end{align}
where $ \AABJM= \AABJM(k,b,\bm\Delta)$. Equation \eqref{eq:ABJMBarnes} is our closed analytic result for the
ABJM Airy constant at general $k$, squashing and chemical potentials.
For $k=1$, $\zeta_6^{(0)}$ reduces to the ordinary six-fold Barnes zeta.

A simple exact check is provided by the round superconformal point,
$b=1$, $\Delta_I=\frac12$. Here the full Molien sum simplifies before
regularization. Writing the character in the form
\begin{align} \label{eq:ABJMroundBarnes}
 \mathcal K_k(t)
 &=\frac{\coth(kt/4)}{2\sinh^2(t/2)}
 \nn\\
 &=\frac{2\mathrm e^{-t}}{(1-\mathrm e^{-t})^2}
 +\frac{4\mathrm e^{-(1+k/2)t}}
 {(1-\mathrm e^{-t})^2(1-\mathrm e^{-kt/2})}\,,
\end{align}
the Mellin transform immediately gives
\begin{align}\label{ABJMBarneszeta}
 \zeta_{\mathcal K_k}(s)
 &=2\zeta(s-1)
 +4\zeta_3\Big(
 s,1+\frac{k}{2}\,\Big|\,1,1,\frac{k}{2}\Big)\,,
\end{align}
where $\zeta(s)$ is the Riemann zeta function and $\zeta_3$ is the Barnes triple zeta.
Computing $\AABJM =\zeta_{\mathcal K_k}'(0)$ then gives the standard
ABJM constant map function $A$:
\begin{align}
 A(k)& \equiv
 \AABJM(k,1,\bm\Delta_{\rm sc}) \nn\\& =
 \frac{2\zeta(3)}{\pi^2k}
 \left(1-\frac{k^3}{16}\right)
 +\frac{k^2}{\pi^2}
 \int_0^\infty\dd x\,
 \frac{x\log(1-\mathrm e^{-2x})}
 {\mathrm e^{kx}-1}\, .
 \label{eq:ABJMconstantmap}
\end{align}
Note that \eqref{ABJMBarneszeta} provides a new Barnes-zeta representation of $A(k)$. 
The result is in agreement with the exact ABJM matrix model and Fermi gas results
\cite{Fuji:2011km,Marino:2011eh}.

Further specializations of \eqref{eq:ABJMBarnes} provide additional
checks against field theory results and are included in Appendix~\ref{app:ABJM}. In Appendix~\ref{app:mABJM} we also determine $\mathcal A$ for mABJM theory \cite{Benna:2008zy, Klebanov:2008vq} and compare
with a new Fermi gas computation.

\section{The large \texorpdfstring{$k$}{k} Type IIA limit}

The large $k$ expansion of the ABJM result of the previous section provides a direct connection
with perturbative Type IIA string theory. In the standard Type IIA limit
one takes $N,k\to\infty$ with the 't Hooft coupling $\lambda=N/k$ fixed,
so that the M-theory circle becomes small and $1/k$ is the genus counting
parameter. The corresponding genus expansion of the ABJM matrix model,
and its relation to topological strings, has been studied extensively starting with
\cite{Marino:2009jd,Drukker:2010nc,Fuji:2011km,Marino:2011eh}.

For the Airy coefficients in \eqref{eq:ABJMCB}, we have
$\cC= O(k^{-1})$ and $\cB=
\frac{k}{24}+O(k^{-1})$.
Thus $\cB$ and $\cC$, which are determined by the localized M-theory action $I_{\rm SUSY}$, 
generate through the Airy function a large $k$ expansion of the standard Type IIA genus form
$F_{\rm pert}^{(\mathcal B, \mathcal C)}=\sum_{g\geq 0}k^{2-2g}f_g(\lambda)$,
with non-trivial functions $f_g(\lambda)$ of the 't Hooft coupling.

By contrast, $\AABJM$ is independent
of the M2-brane chemical potential $\mu$ and contributes a
$\lambda$-independent term at each genus. 
For example, at the round superconformal point, $\AABJM$ reduces to the familiar ABJM constant map function. More generally, its large $k$ structure follows directly from the exact Molien character \eqref{eq:ABJMMolien}. For each $I=1,2,3,4$, let $J$ denote the unique other index with the same $\mathbb Z_k$ charge as $I$, where recall that the orbifold charges are $c=(1,1,-1,-1)$, and let $R,S$ denote the two indices with the opposite charge. 
For each $I$, the reduced weights consist of one difference, $d_I$, involving the other coordinate with the same 
$U(1)_{\rm M}$ charge, and two sums, $r_I,s_I$, involving the coordinates with opposite charge.
We thus define
\begin{align}
d_I
&=
\frac{Q}{2}(\Delta_J-\Delta_I)\,,\quad
r_I
=
\frac{Q}{2}(\Delta_I+\Delta_R)\,,
\quad
s_I
=
\frac{Q}{2}(\Delta_I+\Delta_S)\,,
\label{eq:IIAweights}
\end{align}
so that $d_I+r_I+s_I=Q$. A partial fraction
decomposition followed by the root-of-unity sum gives the exact identity
\begin{align}
\mathcal K_k(t)
&=
\frac12\sum_{I=1}^4
h_I(t)\,
\coth\left(\frac{k Q \Delta_I t}{4}\right)\,,
\label{eq:IIAcharacter}
\end{align}
where
\begin{align}
h_I(t)=\frac{
\sinh(bt)+\sinh(b^{-1}t)
-\sinh(d_I t)-\sinh(r_I t)-\sinh(s_I t)
}{
16\sinh(bt/2)\sinh(b^{-1}t/2)
\sinh(d_I t/2)\sinh(r_I t/2)\sinh(s_I t/2)
}\,.
\label{eq:IIAhI}
\end{align}
This formula is exact at finite $k$; the Type IIA expansion enters only
when the $\coth$ in \eqref{eq:IIAcharacter} is expanded at large $k$. Writing
\begin{align}
h_I(t)
& =
\frac{h_{I,-2}}{t^2}
-\frac{1}{12}
+\sum_{m\geq1}h_{I,2m}t^{2m}\,,
\end{align}
and substituting into the Mellin transform \eqref{eq:zeta}, we obtain
\begin{align}
\AABJM(k,b,\bm\Delta)
&=
A_0(b,\bm\Delta)k^2
-\frac16\log k
+A_1(b,\bm\Delta)
+\sum_{g\geq2}A_g(b,\bm\Delta)k^{2-2g}\,,
\label{eq:IIAlargek}
\end{align}
where, for $g\geq2$,
\begin{align}
A_g(b,\bm\Delta)
&=
(2g-3)!\,\zeta(2g-2)
\sum_{I=1}^4
\frac{h_{I,2g-2}}{(Q\Delta_I/2)^{2g-2}}\,.
\label{eq:IIAAg}
\end{align}
The term $h_{I,-2}$ gives the genus zero contribution, while the universal
constant $-\frac{1}{12}$ gives the $-\frac{1}{6}\log k$ term, and a contribution to the finite constant $A_1$. The remaining contribution to $A_1$ is different in character, since it depends on the full
zeta-regularized determinant rather than on a single coefficient in the
small $t$ expansion.

The dependence on the squashing parameter can be organized further. Rescaling the reduced weights by $Q=b+b^{-1}$, 
one finds the all-genus finite refinement grading
\begin{align}
A_g(b,\bm\Delta)
&=
\sum_{n=0}^{g}
Q^{2n-2g+2}\,
\cU_{g,n}(\bm\Delta)\,,
\qquad g\geq2\,,
\label{eq:IIAgrading}
\end{align}
where the coefficients $\cU_{g,n}$ depend only on the chemical
potentials $\bm\Delta$.
Thus $Q^{2g-2}A_g$ is a polynomial of degree at most $g$ in $Q^2$.
This is the standard finite grading obtained from the refined
topological string expansion at fixed total genus.

To make the Type IIA interpretation explicit, introduce the reduced
genus-counting parameter
\begin{align}
g_{\rm red}
&\equiv 
\frac{4\pi\ii}{kQ}\,.
\label{eq:IIAgred}
\end{align}
At $b=1$, for which $Q=2$, this reduces to the usual ABJM parameter
$2\pi\ii/k$. The contribution of $\AABJM$ to the Type IIA free energy
may then be written as
\begin{align}
-\sum_{g\geq2}k^{2-2g}A_g
&=
\sum_{g\geq2}\sum_{n=0}^{g}
(-1)^n g_{\rm red}^{2g-2}Q^{2n}
\widehat F^{\rm red}_{g,n}(\bm\Delta)\,,
\label{eq:IIAFred}
\end{align}
with
\begin{align}
\widehat F^{\rm red}_{g,n}
&=
\frac{(-1)^{n+1}}{(4\pi\ii)^{2g-2}}\,
\cU_{g,n}\,.
\end{align}
In particular, the $n=0$ component of this refinement grading takes the remarkably simple all-genus form
\begin{align}
\widehat F^{\rm red}_{g,0}
&=
\kappa_g\,\mathcal E^{\rm red}_g(\bm\Delta)\,,
\qquad g\geq2\,,
\nn\\
\kappa_g
&=
\frac{(-1)^g}{2(2g-2)!}
\frac{|B_{2g}|}{2g}
\frac{|B_{2g-2}|}{2g-2}\,,
\label{eq:IIAHodge}
\end{align}
where $B_{2g}$ is the $2g^{\rm th}$ Bernoulli number.
The coefficient $\kappa_g$ is precisely the universal constant-map coefficient of the degree-zero, higher-genus Gromov--Witten free energy
 \cite{FaberPandharipande}. Thus the $n=0$ component of the eleven-dimensional one-loop determinant reproduces the standard worldsheet constant map coefficient for every $g\geq2$, while also determining the non-trivial equivariant factor $\mathcal E^{\rm red}_g(\bm\Delta)$. The remaining components $\cU_{g,n}$, $n>0$, give its finite refined completion. Explicitly,
\begin{align}
\mathcal E^{\rm red}_g(\bm\Delta)
=
\sum_{I=1}^4
\Delta_I^{\,2-2g}
\frac{(d_I+r_I)(d_I+s_I)(r_I+s_I)}
     {d_Ir_Is_I}\,.
\end{align}

This equivariant factor also has a direct geometric interpretation. Promoting the $\mathbb Z_k$ action in \eqref{eq:ABJMMolien} to its parent $U(1)$ and rewriting the Molien sum as a contour integral, the internal residues remove the M-theory circle direction from both the tangent denominator and the $T-T^*$ numerator. The resulting character is the tangent-cotangent character of the reduced complex three-dimensional geometry, while the line bundle associated with the M-theory circle is determined by the RR two-form flux in Type IIA. In Appendix~\ref{sec:IIAreduction} we give this circle reduction construction and show that the expression above is a flux-dependent equivariant characteristic number of the reduced geometry.

\section{Topologically twisted index}

We next consider the refined topologically twisted index (TTI) of ABJM on
$S^1\times S^2$ \cite{Benini:2015noa}, which counts microstates of
magnetically charged, rotating $AdS_4$ black holes
\cite{Benini:2015eyy,Bobev:2022jte,Bobev:2023lkx}. As in~\cite{BenettiGenolini:2026cyc}, the corresponding bulk saddle has
$M_4\simeq\mathbb R^2\times S^2$, with $S^2$ the Euclidean horizon at the
origin of $\mathbb R^2$. We normalize the supersymmetric Killing vector $K$
to have unit weight in the $\mathbb R^2$ plane. It generates a $U(1)$
rotation of the horizon and has two fixed points, at the north and south
poles of $S^2$, which we label by $\sigma=\pm$. The corresponding tangent
weights along the horizon are $\sigma\epsilon$, where $\epsilon$ is the
refinement/rotation parameter.

At each pole there are two external and four internal holomorphic tangent
weights. With the normalization above, the external weights at the fixed point
$\sigma=\pm$ are
\begin{align}
	b_{\sigma,1}=1\,,
	\qquad
	b_{\sigma,2}=\sigma\epsilon\, .
\end{align}
The four internal weights are determined by the ABJM chemical potentials
$\Delta_I$ and magnetic fluxes $\mathfrak n_I$, which specify the twisting of
the four internal complex directions of the fibre in \eqref{fibredform} over
the horizon $S^2$. Using the same bulk/boundary identification as in \eqref{eq:ABJMvariables}, they are
\begin{align}
u_{\sigma,I}
=
-\frac{1}{2}
\left(\Delta_I+\sigma\epsilon\mathfrak n_I\right)\,.
\end{align}
Thus the six weights at the pole $\sigma=\pm$ are
\begin{align}
	w_\sigma
	&=
	\Big(
	1,\sigma\epsilon,
	-\frac{\Delta_1+\sigma\epsilon\mathfrak n_1}{2},
	\ldots,
	-\frac{\Delta_4+\sigma\epsilon\mathfrak n_4}{2}
	\Big)\, .
	\label{eq:TTIweights}
\end{align}
The condition \eqref{eq:CYweights} that the Killing spinor is uncharged under
$K$ must hold at both fixed points. Using \eqref{eq:TTIweights} gives
\begin{align}
	\sum_\alpha w_{\sigma,\alpha}
	=
	1+\sigma\epsilon
	-\frac{1}{2}
	\Big(
	\sum_I\Delta_I
	+\sigma\epsilon\sum_I\mathfrak n_I
	\Big)
	=0\, ,
\end{align}
and hence
\begin{align}
	\sum_I\Delta_I=2\,,
	\qquad
	\sum_I\mathfrak n_I=2\, .
	\label{eq:TTIconstraints}
\end{align}
The first equation is the usual ABJM chemical potential constraint, while the second
is the topological twist on~$S^2$.
For a globally well-defined eleven-dimensional uplift there is in addition a quantization condition on the magnetic fluxes. 
Since the internal fibre is
 $\C^4/\Z_k$, the allowed twists take values in the co-character lattice of  
$U(1)^4/\Z_k$. For the orbifold charges
$c=(1,1,-1,-1)$ this gives
\begin{align}
	\boldsymbol{\mathfrak n}
	\in
	\Z^4+\frac{1}{k}(1,1,-1,-1)\Z\, .
	\label{eq:TTIfluxlattice}
\end{align}
In particular, the universal $S^2$ twist
$\mathfrak n_I=\frac{1}{2}$ is globally well-defined precisely when $k$ is even.

The local equivariant $G$-flux parameters \eqref{eq:mudef} associated with the fixed points $(\sigma,a)$ of the external and internal geometries are labelled $\mu_{\sigma,a}$. 
As in the $EAdS_4$ backgrounds discussed above \eqref{eq:Zpert_Jpert}, we restrict to configurations in which they are independent of the internal fixed point, $\mu_{\sigma,a} \equiv \mu_\sigma$. Using the weights \eqref{eq:TTIweights} in \eqref{eq:ISUSYfixed}, 
we then obtain
\begin{align}
	I_{\rm SUSY} &= - \sum_\sigma \Big( \frac{\cC_\sigma}{3} \, \mu_\sigma^3 + \cB_\sigma \, \mu_\sigma \Big) \, ,
	\end{align}
	where 
	\begin{align}
	\cB_\sigma &= \frac{1}{48 k \, \prod_I (\Delta_I + \sigma \epsilon \mathfrak{n}_I) } \Bigg[ - 4 (1-\epsilon^2)^2 
	+ \frac{1}{2} \sum_{I<J} (\Delta_J+\sigma  \epsilon  \mf{n}_J)^2 (\Delta_I+\sigma  \epsilon  \mf{n}_I)^2 \nn \\
	& \qquad\qquad  + 2 \left( 1 + \epsilon^2 \right) \sum_I (\Delta_I+\sigma  \epsilon  \mf{n}_I)^2 
	- \frac{1}{4}  \sum_I (\Delta_I+\sigma  \epsilon  \mf{n}_I)^4 \Bigg] + \frac{k^2-1}{24 k} \, , \nn \\
	\cC_\sigma &= \frac{2 \epsilon^2 }{\pi^2 k \prod_I (\Delta_I + \sigma \epsilon \mf{n}_I) } \, .
\end{align}

The two remaining local flux parameters $\mu_\pm$ obey, by
\eqref{eq:mutotal} and \eqref{eq:diagonalmu},
\begin{align}
 \mu_++\mu_-=\mu\, .
\end{align}
Since the total chemical potential $\mu$ is fixed as part of the conformal boundary data $\mathscr{B}$, the kernel
\eqref{eq:mukernel} is one-dimensional,
$\delta\mu_++\delta\mu_-=0$. We parametrize this relative flux degree of
freedom by
\begin{align}
 \mu_+=\nu\,,
 \qquad
 \mu_-=\mu-\nu\,.
\end{align}
Thus $\nu$ is precisely the residual localized variable integrated over in
\eqref{eq:master_2}. 

The measure $[\dd\nu]$ in \eqref{eq:master_2} was
left schematic. In the present case we propose to take
$[\dd\nu]=\dd\nu/(2\pi\ii)$, with the integration cycle specified separately.
Denoting the cycle at fixed $\mu$ by $\Gamma_\nu(\mu)$, we have
\begin{align}
Z^{\rm pert}_{\rm M}[\mu]
&=
Z_{\text{1-loop}}
\int_{\Gamma_\nu(\mu)}
\frac{\dd\nu}{2\pi\ii} \, \exp\left[-I_{\rm SUSY}(\mu,\nu)\right]\,.
\label{eq:TTIgrandcanonical}
\end{align}
Here $Z_{\text{1-loop}}$ is independent of $\mu$ and $\nu$, since the
one-loop character depends only on the equivariant tangent weights and not
on the independent equivariant $G$-flux parameters.
To obtain the TTI at fixed $N$, one then performs the inverse Laplace transform in $\mu$. We choose the combined integration cycle 
so that, in the $\mu_\pm$ variables it is represented by the product of the block cycles $\Gamma_\pm$. With this prescription, the canonical partition function factorizes as
\begin{align}
Z_{\rm M}^{\rm pert}[N]
&=
Z_{\text{1-loop}}
\prod_{\sigma=\pm}
\int_{\Gamma_\sigma}
\frac{\dd\mu_\sigma}{2\pi\ii}
\exp\left[
-I_{\rm SUSY}^{\sigma}(\mu_\sigma)
-N\mu_\sigma
\right]\,.
\label{eq:TTIcanonical}
\end{align}
The
result is the product of the two Airy gravitational blocks discussed in
\cite{BenettiGenolini:2026cyc,Hristov:2022lcw,Bobev:2026lvl}.
The cycles $\Gamma_\sigma$, including their analytic continuation in the equivariant parameters, are part of the prescription. 
As shown in Appendix~\ref{sec:TTIunrefined}, the unrefined
limit $\epsilon\to0$ reproduces the known field theory
structure when the cycles $\Gamma_\sigma$ are chosen so
that the perturbative saddle point expansions of the
block integrals are taken about
\begin{align}\label{saddassump}
 \mu_\sigma^*
 =
 \sigma\sqrt{\frac{N-\cB_\sigma}{\cC_\sigma}}\, .
\end{align}

It remains to determine the one-loop factor. 
 Up to a common rescaling,
\eqref{eq:TTIweights} has the ABJM form \eqref{eq:ABJMnormalizedweights},
with
\begin{align}
b_\sigma=(\sigma\epsilon)^{-1/2}\,,
\qquad
\Delta_{\sigma,I}
=\frac{\Delta_I+\sigma\epsilon\mathfrak n_I}
{1+\sigma\epsilon}\,.
\label{eq:TTIeffective}
\end{align}
Since $\zeta_{\mathcal K_\sigma}(0)=0$, the common rescaling drops
out and
\begin{align}
\log Z_{\text{1-loop}}
&=
\sum_{\sigma}
\AABJM\!\left(k,b_\sigma,\bm\Delta_\sigma\right)  = \cA_+ + \cA_- \,,
\label{eq:TTIA}
\end{align}
where 
\begin{align} 
\cA_+  = & \ \frac{1}{k}\partial_s\sum_{r=0}^{k-1}\Big\{
\zeta_6^{(r)}(s,1)
-\zeta_6^{(r)}(s,1+2\epsilon)
+\zeta_6^{(r)}(s,\epsilon)
-\zeta_6^{(r)}(s,2+\epsilon)
\nonumber\\
& \quad +\sum_{I=1}^{4}\omega^{-r c_{I+2}}
\zeta_6^{(r)}\!\Big(
s,\Big(1+\epsilon+\frac{\Delta_I+\epsilon\mf{n}_I}{2}\Big)\Big)
\nonumber\\
& \quad
-\sum_{I=1}^{4}\omega^{r c_{I+2}}
\zeta_6^{(r)}\!\Big(
s,\Big(1+\epsilon-\frac{\Delta_I+\epsilon\mf{n}_I}{2}\Big)\Big)
\Bigr\}\Big|_{s=0}\, .
\label{eq:TTIBarnes}
\end{align}
Here the twisted Barnes zeta function \eqref{eq:twistedBarnes} has periods
\begin{equation}
 \bm\Omega_6^+\equiv
 \Big(
 1,\epsilon,
 \frac{\Delta_1+\epsilon \mf{n}_1}{2},\ldots,
 \frac{\Delta_4+\epsilon \mf{n}_4}{2}
 \Big)\, .
 \label{eq:TTIOmega}
\end{equation}
The second block $\cA_-$ is understood by analytic continuation in the equivariant weights.
Thus the refined TTI determinant is expressed in terms of the same twisted Barnes-zeta/Molien construction as the sphere partition function.

 The ordinary unrefined TTI is obtained by taking the smooth $\epsilon\to0$ limit of the full result 
\eqref{eq:TTIcanonical}, using the saddle choices in \eqref{saddassump}. 
This limit is singular at the
level of an individual fixed point character and must therefore be taken
after zeta regularization and the Airy block integrals. As shown in
Appendix~\ref{sec:TTIunrefined}, the result takes the all-orders form 
\begin{align}
 \log Z_{\rm M}^{\rm pert}[N]
 &=
 -\frac{\pi\sqrt{2k\prod_I\Delta_I}}{3}
 \sum_I\frac{\mathfrak n_I}{\Delta_I}
 \left(
 \widehat N_{k,\Delta}^{3/2}
 -\frac{\cnew_I}{k}\widehat N_{k,\Delta}^{1/2}
 \right)
 \nonumber\\
 &\qquad
 -\frac{1}{2}\log\widehat N_{k,\Delta}
 +\widehat f_0(k,\bm\Delta,\bm{\mf{n}})\,,
 \label{eq:TTIallorders}
\end{align}
where
$\widehat N_{k,\Delta}
=N-k/24+(12k)^{-1}\sum_I\Delta_I^{-1}$ and
$\cnew_I=\cnew_I(\bm\Delta)$ is given in~\cite{Bobev:2022jte}:
\begin{equation}
	\cnew_I \equiv \frac{\prod_{J\neq I}(\Delta_I + \Delta_J)}{8 \Delta_1 \Delta_2 \Delta_3 \Delta_4} \sum_{J\neq I} \Delta_J \, .
\end{equation}
The first terms reproduce the known all-orders field theory structure,
while the one-loop determinant determines the $N$-independent constant
$\widehat f_0(k,\bm\Delta,\bm{\mf{n}})$. A closed form expression for
generic chemical potentials and magnetic charges is derived in
Appendix~\ref{sec:TTIunrefined}.
At the superconformal point
$\Delta^{\rm sc}_I=\mathfrak n^{\rm sc}_I=\frac{1}{2}$, this simplifies to
\begin{align}
 \widehat f_0(k,\bm\Delta^{\rm sc},\bm{\mf{n}}^{\rm sc})
 &=
 -4A(k)-3k\,A(4/k)
 +\frac{3k}{2}\,A(8/k)
 -\frac{k^2}{2}A(2)\nonumber\\
& \quad \ -\frac{1}{2}\log k
 -\frac{5}{2}\log 2\,,
 \label{eq:TTIf0}
\end{align}
in agreement with equation (42) of \cite{Hong:2026zul}
(see also \cite{Hosseini:2026lvj}).

\section{\texorpdfstring{$V^{5,2}$}{V52} theory}

We next apply the prescription of section \ref{locpisec} to the squashed $S^3_b$ partition
function of the $\mathcal N=2$ SCFT \cite{Martelli:2009ga} dual to M-theory
on $AdS_4\times V^{5,2}$, where $V^{5,2}=SO(5)/SO(3)$ is a homogeneous Sasaki--Einstein manifold. The latter is non-toric, and hence lies outside the class of toric
Calabi--Yau four-fold cones considered in~\cite{BenettiGenolini:2026cyc}.
The twelve-dimensional filling is
$W\simeq \C^2\times X$, where the Calabi--Yau cone is the quadric
hypersurface
\begin{equation}
 X=C(V^{5,2})
 =\left\{z_1z_2+z_3z_4+z_5^2=0\right\}
 \subset\C^5\, .
 \label{eq:V52cone}
\end{equation}
The apex of $X$ is a four-dimensional ordinary double point singularity. 
Unlike the ordinary double point at the apex of the three-fold conifold, which admits two small crepant resolutions, this point admits no crepant resolution: the singularity
is factorial and terminal, and in particular the ordinary blow-up is
non-crepant. The standard smoothing deformation, which is diffeomorphic to $T^*S^4$,
breaks the Reeb $\C^*$ action and hence is not an equivariant filling
for the localization problem. However, it is straightforward to compute 
everything directly on the cone, as we now explain. 

As in the ABJM analysis of section~\ref{secabjm}, we normalize the external weights to $b,b^{-1}$ and write $Q=b+b^{-1}$. 
The two Cartan equivariant parameters $m_1,m_2$ for the Cartan of the $SO(5)$ flavour symmetry,\footnote{The real mass parameters of the dual field theory are
proportional to $\ii m_i$.} shift the five ambient internal weights. Denoting by $w_P$ the weight of the defining quadric in \eqref{eq:V52cone}, we then have
\begin{align}
\left(w_1,\ldots,w_5;w_P\right)
&=
\left(
-\frac{Q}{3}-m_1,
-\frac{Q}{3}+m_1,
-\frac{Q}{3}-m_2,
-\frac{Q}{3}+m_2,
-\frac{Q}{3};
-\frac{2Q}{3}
\right)\,.
\label{eq:V52weights}
\end{align}
Since $X$ is a hypersurface, the relevant fixed point data are encoded by the intrinsic virtual tangent class
$T_X^{\rm vir}=T\C^5|_X-N_X$.\footnote{See Appendix~\ref{sec:V52appendix} for a discussion of the intrinsic virtual class and its behaviour under equivariant resolutions.}
Including the two external directions, its total equivariant weight is
$Q-5Q/3+2Q/3=0$, as required by \eqref{eq:zerochargesplit}.

As for the ABJM $S^3_b$ partition function we have a single 
M2-brane chemical potential $\mu$, with no relative
$\nu$-integration in \eqref{eq:master_2}. Evaluating \eqref{eq:ISUSYfixed} on the corresponding virtual fixed point data gives 
\begin{align}
 \cB&=
 \frac{1}{12(Q^2-9m_1^2)(Q^2-9m_2^2)}
 \Bigl[-5Q^4+36Q^2
 +(18Q^2-81)(m_1^2+m_2^2)+81m_1^2m_2^2\Bigr] ,
 \nonumber\\
 \cC&=
 \frac{81}
 {4\pi^2(Q^2-9m_1^2)(Q^2-9m_2^2)}\, .
 \label{eq:V52CB}
\end{align}
Setting $m_1=m_2=0$ agrees with the field theory
results of \cite{Bobev:2025ltz} for the superconformal R-symmetry; the general result \eqref{eq:V52CB} is new and in particular gives a new prediction for field theory. 
One can derive the same formulas \eqref{eq:V52CB} using the non-crepant 
blow-up  $\widetilde X=\operatorname{Tot}\!\left(\mathcal O_{\mathcal{Q}^3}(-1)\right)$,
where the complex quadric three-fold
$\mathcal{Q}^3\simeq SO(5)/(SO(2)\times SO(3))$ is the exceptional divisor.

For the one-loop character it is convenient, as in the ABJM calculation,
to rewrite the negative internal weights using decaying exponentials. Define 
\begin{align}
 q_1&=\mathrm e^{-bt}\, , &\ q_2 &=\mathrm e^{-b^{-1}t}\, ,
& \ z =\mathrm e^{-2Qt/3}\, , \quad \,  x_5=\mathrm e^{-Qt/3}\, ,
 \nonumber\\
 x_{1,2}&=\mathrm e^{-(Q/3\pm m_1)t}\, , &\
 x_{3,4}&=\mathrm e^{-(Q/3\pm m_2)t}
\, .
 \label{eq:V52multweights}
\end{align}
The hypersurface analogue of \eqref{eq:character} is then
\begin{align}
&  \mathcal K_{V^{5,2}}(t)
 =\frac{q_1q_2(1-z)}
 {(1-q_1)(1-q_2)\prod_{i=1}^5(1-x_i)}
 \Bigg\{
 \sum_{\alpha=1}^2(q_\alpha^{-1}-q_\alpha)
 +\sum_{i=1}^5(x_i-x_i^{-1})+z^{-1}-z\Bigg\}.
 \label{eq:V52char}
\end{align}
Here $(1-z)$ imposes the quadratic hypersurface relation in the Dolbeault
character, while $z^{-1}-z$ accounts for the normal direction in the
virtual tangent-cotangent class. As for ABJM, the localized action has no
$\mu$-independent term in this sphere sector, and hence
\begin{equation}
 \cA_{V^{5,2}}(b,m_1,m_2)
 =\zeta'_{\mathcal K_{V^{5,2}}}(0)\, .
 \label{eq:V52Agen}
\end{equation}
This is a closed analytic result at general squashing and
equivariant masses.

At the superconformal point $m_1=m_2=0$, the character reduces to a finite combination of ordinary six-fold Barnes zeta functions. In the conventions of~\cite{Bobev:2025ltz}, in the small $b$ limit
we define $\widehat g_0^{V^{5,2}}$ by
\begin{align}
\cA_{V^{5,2}}(b,0,0)
&=
-\frac{\widehat g_0^{V^{5,2}}}{b^2}
+O(\log b)\,,
\qquad b\to0\,.
\end{align}
As shown in Appendix~\ref{sec:V52appendix}, we find
\begin{align}
\widehat g_0^{V^{5,2}}
&=
\frac{20}{9}\zeta'(-2)
-\frac{2}{9}L'(-1,\chi_{-3})\nn\\
&=
-0.139455670622974\ldots\,,
\label{eq:V52g0}
\end{align}
where $\chi_{-3}$ is the primitive Dirichlet character modulo three,
in agreement with the independent numerical field theory result of~\cite{Bobev:2025ltz}.

The above construction may be extended to the
$\Z_k$ quotient $V^{5,2}/\Z_k$ \cite{Martelli:2009ga}. In the
coordinates of \eqref{eq:V52cone} here the $\Z_k\subset U(1)_b$ action has charges	$c=(1,-1,1,-1,0)$
on $(z_1,\ldots,z_5)$.\footnote{This $\Z_k$ quotient is the
Martelli--Sparks quotient associated with the Chern--Simons level \cite{Martelli:2009ga}, and
should not be confused with the $\Z_{N_f}$ quotient appearing in the
mirror flavoured Jafferis description of the $V^{5,2}$ theory \cite{Jafferis:2009th}, which arises
from a different $U(1)\subset SO(5)$ action with a fixed locus.} In \eqref{eq:V52CB} the leading coefficient $\cC$ is consequently
divided by $k$, 
while $\cB$ is not simply rescaled by $1/k$ but also receives an orbifold contribution. 
The one-loop determinant is obtained
by the same Molien projection as in the ABJM example, namely by averaging
\eqref{eq:V52char} over $x_i\to\omega^{r c_i}x_i$, with
$\omega=\mathrm e^{2\pi \ii/k}$. Since this quotient is along the
$U(1)_b$ M-theory circle, its large $k$ expansion likewise has a
Type IIA interpretation.
More generally, other Calabi--Yau four-fold cones and their discrete
quotients may be treated in the same way, using their equivariant
 data in the localized action and one-loop character.

\section{Discussion}
\label{sec:Discussion}

The results of this work provide evidence for a protected perturbative
localization of M-theory based on relative equivariant localization of
eleven-dimensional supergravity. The localized action reproduces
the M2-brane Airy coefficients $\cB$ and $\cC$, while the proposed one-loop
fluctuation character determines $\cA$ by zeta regularization. 
The agreement with exact ABJM and mABJM results, together with the non-toric $V^{5,2}$ example, provides non-trivial tests of a common geometric framework. The general formulas also give predictions in regimes where independent field theory calculations are not yet available.

Our prescription for the one-loop term in \eqref{eq:master_2} remains a
conjecture for the protected M-theory path integral. Its tangent-cotangent
character is motivated by existing constructions
\cite{Raghavendran:2021qbh,Hahner:2023kts,Costello:2018zrm,Nekrasov:2014nea},
but these inputs do not by themselves establish the relative
twelve-dimensional character as the physical one-loop determinant of the
holographic saddles considered here. A first-principles derivation should
identify the gauge-fixed supersymmetry complex and reproduce both the
character and its determinant normalization.

 In view of the new structure uncovered here, it would be interesting to revisit earlier direct eleven-dimensional one-loop calculations.
 The universal logarithmic correction to
the ABJM sphere free energy was reproduced in \cite{Bhattacharyya:2012ye}
from a normalizable two-form ghost zero mode, while the subsequent attempt
of \cite{Liu:2016rdi} to determine the finite $N^0$ term from the
Kaluza--Klein spectrum did not reproduce the correct $k$ dependence. A
recent gauge-fixed re-analysis \cite{Arrighi:2026kk} has emphasized that
ghosts, $AdS$ modes and the resummation of the full KK tower already
interact non-trivially in the logarithmic term. The character obtained here
provides a concrete formula for extending these calculations to the finite
part: a direct derivation should incorporate the complete supersymmetric
fluctuation complex, the boundary conditions appropriate to the fixed $C$-field
ensemble \cite{Bobev:2026gir}, and a regulator acting on the full KK tower
rather than on its separate components.
Related localization constructions developed in different settings may also be worth revisiting in this context, 
including the index-theoretic derivation of factorized one-loop logarithmic corrections in $AdS_4$ supergravity \cite{Hristov:2021zai}, as well as \cite{Dabholkar:2010uh,Iliesiu:2022kny,Dabholkar:2014wpa,Costello:2016nkh,Gaiotto:2019wcc,Gaiotto:2020vqj,DelZotto:2021gzy}. In particular, it would be interesting to understand whether ingredients from these approaches, together with the treatment of asymptotic boundaries \cite{deWit:2018dix} and equivariant localization of supergravity actions \cite{BenettiGenolini:2023kxp,BenettiGenolini:2026qdm,BenettiGenolini:2026cdw}, can help in deriving the relevant quantum complex.

The formalism is considerably more general than the examples studied here.
For example, suppose that the Calabi--Yau four-fold cone $X=C(Y_7)$ admits
an equivariant toric crepant resolution $\widetilde X$ (which is smooth) with isolated fixed
points $a\in\widetilde X$. Let $u_{a,I}$, $I=1,\ldots,4$, denote
the four toric weights on $T_a\widetilde X$, and let $b_1,b_2$ denote the two
external weights. Defining
\begin{align}
 q_i&=\mathrm e^{-t b_i}\,,\qquad
 q_{a,I}=\mathrm e^{-t u_{a,I}}\,,
\end{align}
with
$ b_1+b_2+\sum_{I=1}^4u_{a,I}=0$, 
the character \eqref{eq:character-general} reduces to the fixed point sum
\begin{align}
 \mathcal K_W(t)
 &=
 \sum_{a\in\widetilde X}
 \frac{
 (q_1^{-1}-q_1)+(q_2^{-1}-q_2)
 +\sum_{I=1}^4(q_{a,I}^{-1}-q_{a,I})
 }{
 (1-q_1)(1-q_2)\prod_{I=1}^4(1-q_{a,I})
 }\, .
 \label{eq:toric-general-character}
\end{align}
In particular, for three-sphere partition functions the one-loop contribution to
the Airy constant follows directly from the toric fixed point data through
$\cA=\zeta'_{\mathcal K}(0)$. 
As discussed in section~\ref{locpisec}, \eqref{eq:toric-general-character} is independent
of the choice of crepant toric  resolution $\widetilde{X}$.\footnote{The ABJM example is singular, with internal space
$\C^4/\Z_k$, and is treated directly using the Molien projection
in \eqref{eq:ABJMMolien}. For more general orbifold singularities
the corresponding one-loop character may require additional
twisted sector contributions. We leave a systematic treatment of
such singular backgrounds for future work.}
More generally, the equivariant $C$-field data may incorporate discrete
torsion $G$-flux, with non-trivial flux through the filling, and baryonic
chemical potentials, encoded in the remaining fixed point parameters
$\mu_A$. These more general sectors, and the associated global constraints
on the equivariant flux data, will be developed elsewhere.

The same framework should also extend naturally to topologically twisted
indices on $S^1\times\Sigma_g$. For the corresponding
higher-genus saddles the supersymmetric Killing vector has
non-isolated fixed loci in the bulk, corresponding to $\Sigma_g$ bolt
components rather than the isolated fixed points encountered for $S^2$.
Both \eqref{ISUSYloc} and \eqref{eq:character-general} were written for a
general fixed locus $\mathscr F$, and in this case the corresponding
equivariant characteristic classes are integrated over the bolt. This
provides the natural extension of the localized action and one-loop
determinant to such saddles.

The dependence of the one-loop determinant on equivariant flavour parameters encodes interesting information on integrated correlators of the undeformed
theories. For example, in ABJM
theory the generic chemical potential dependence determines integrated
four-point correlators of the type studied in
\cite{Binder:2019mpb,Chester:2021gdw}.
As discussed in Appendix~\ref{app:correlate}, one such combination probes the additional three-mass information of generic $\mathcal N=6$ 
ABJM and departs from the $\mathcal N=8$ relation, valid for $k=1,2$, only at order $N^0$, corresponding holographically to a one-loop bulk correction.

Our strategy is to combine supersymmetry and equivariant localization
with precise field theory input, using holography to constrain ingredients
of the bulk prescription that are not fixed by local geometry alone. In
particular, while the localized action and one-loop character determine the
integrand in \eqref{eq:master_2}, they do not in general determine the
measure $[\dd\nu]$ or the integration cycle over the remaining relative
equivariant data. 
The topologically twisted index provides one example in
which field theory information constrains these choices. Picard--Lefschetz
theory may also provide a useful framework for determining the appropriate
integration cycles and their decomposition into thimbles.

Much about the full M-theory path integral remains to be understood. Even within the class of backgrounds studied here, important open problems include the global determinant line structure and $\eta$-invariant phases, and the inclusion of additional supersymmetric saddles and non-perturbative sectors. More generally, the holomorphic $TW-T^*W$ character used in this paper relies on a global complex structure compatible with the preserved supersymmetry, and is therefore not expected to apply unchanged to all supersymmetric backgrounds. For example, for saddles relevant to the superconformal index the four-dimensional Killing spinor contains both chiralities, with opposite chiral components vanishing at the two fixed points, suggesting that the corresponding one-loop complex may require a more general geometric formulation, perhaps in terms of generalized or exceptional geometry. Different holographic boundary conditions, including those appropriate to M5-brane backgrounds, provide further important directions and will likewise require modifications of the fluctuation complex. The framework and results developed here open the way to a systematic study of a much wider class of protected observables in quantum M-theory.

\section*{Acknowledgments} 

We thank Shai Chester for helpful discussions.
This work was supported
in part by STFC grants ST/X000575/1 and ST/X000761/1, and also the SNSF Ambizione
grant PZ00P2\_208666. JP is supported by a Dean's PhD studentship at
Imperial College. FG is supported by an STFC studentship.
Some of the ideas and calculations were developed with the assistance of AI tools (ChatGPT and Claude). 

\appendix
\section{Further checks of the ABJM Airy constant \texorpdfstring{$\AABJM$}{AABJM}}\label{app:ABJM}

We can compare our general expression \eqref{eq:ABJMBarnes} for $\AABJM(k,b,\bm\Delta)$ with
field theory results that are known for special loci. It is simplest
to first reduce the character $\mathcal K_k(t)$ in \eqref{eq:ABJMMolien} and then apply the general
zeta-regularized definition in \eqref{eq:zeta}.

\paragraph{Nosaka locus.} 
For general $k$, take
\begin{align}
\label{eq:NosakaLocus}
b&=1\,,\qquad
\Delta_1+\Delta_3=\Delta_2+\Delta_4=1\,.
\end{align}
The character \eqref{eq:ABJMMolien}  decomposes into four copies of the
superconformal character appearing in
\eqref{eq:ABJMroundBarnes}, and hence
\begin{align}
\left.
\AABJM(k,1,\bm\Delta)
\right|_{\rm Nosaka}
&=
\frac14\sum_{I=1}^4
A(2k\Delta_I)\,,
\label{eq:ABJMNosaka}
\end{align}
with the constant map function $A$ defined in \eqref{eq:ABJMconstantmap}.
This agrees with~(1.12) of~\cite{Nosaka:2015iiw}. The closed expression for the constant was
presented there as a conjecture, supported by the Fermi gas WKB expansion
and finite $(k,N)$ data. 
In the $\Delta_I$ conventions used here, the same result is summarized in equations~(2.27)--(2.28) of~\cite{Bobev:2025ltz}, up to a permutation of the $\Delta_I$.

\paragraph{Kubo--Nosaka--Pang locus.}
At $k=1$ we consider the squashed locus
\begin{align}
\Delta_1&=\frac12-\frac{2\ii\zetanew}{Q}\,,\qquad
\Delta_2=\frac{2b^{-1}}{Q}\,,\nn\\
\Delta_3&=\frac12+\frac{2\ii\zetanew}{Q}\,,\qquad
\Delta_4=\frac{b-b^{-1}}{Q}\,.
\end{align}
Using the ABJM tangent weights in \eqref{eq:ABJMnormalizedweights} and exploiting the invariance of
$\zeta'_{\mathcal K}(0)$ under a common rescaling of all weights,
the Airy constant reduces to
\begin{align}
\left.
\AABJM(1,b,\bm\Delta)
\right|_{\rm KNP}
&={}\frac14\Bigg[
A\left(\frac{b^2+1-4\ii b\zetanew}{2}\right)
+A\left(\frac{b^2+1+4\ii b\zetanew}{2}\right)\nn\\
& \qquad \quad+A(b^2-1)-A(2b^2)
\Bigg]\,.
\label{eq:ABJMKNP}
\end{align}
The corresponding Airy constant was obtained in equations~(3.39)--(3.40) in
\cite{Kubo:2024qhq}, from their Fermi gas
construction. Their analytic derivation applies on the special locus
$b^2\in2\mathbb N-1$, with continuation in $b$ supported numerically.
In the $\Delta_I$ conventions used here, the same result is written as equation~(2.33) of~\cite{Bobev:2025ltz}.

The point $b^2=3$, $\zetanew=0$ has $\Delta_I=1/2$ and was solved earlier by
Hatsuda \cite{Hatsuda:2016uqa}.  Equation~\eqref{eq:ABJMKNP} reduces to 
\begin{align}
\left. \AABJM \right|_{\rm Hatsuda}
=
-\frac{\zeta(3)}{3\pi^2}+\frac16\log3\,,
\end{align}
in agreement with~(3.23) of~\cite{Hatsuda:2016uqa}.

\paragraph{Generic large $k$ check.}
The preceding exact checks lie on special loci in parameter space.
As a complementary check of the large $k$ expansion
\eqref{eq:IIAlargek}, we may instead keep $b$ and the $\Delta_I$
arbitrary. The leading term gives
\begin{align}
\AABJM(k,b,\bm\Delta)
&=
-\mathfrak c^{\rm ABJM}(b,\bm\Delta)\,k^2
+o(k^2)\,,
\end{align}
where $\mathfrak c^{\rm ABJM}(b,\bm\Delta)$ agrees term by term with equation~(4.2b) of~\cite{Bobev:2025ltz}. This provides a check of the
genus-zero coefficient $A_0(b,\bm\Delta)$ for completely generic
squashing and chemical potentials. We note, however, that the generic
field theory expression in~\cite{Bobev:2025ltz} was inferred from
exact special loci, symmetries and matrix model constraints, rather than
from a direct saddle point computation at generic $(b,\bm\Delta)$; see
section~4.1.2 of that reference.

\section{The mABJM specialization}\label{app:mABJM}

The mABJM theory is reached by a superpotential mass perturbation of ABJM with $k=1,2$. This breaks $\cN=8$ supersymmetry to $\cN=2$, and triggers an RG flow that terminates in a SCFT with $SU(3)$ flavour symmetry \cite{Benna:2008zy, Klebanov:2008vq}, holographically dual to the $\mathcal N=2$, $SU(3)\times U(1)_R$-invariant $AdS_4$ 
vacuum \cite{Warner:1983vz,Corrado:2001nv}.
In terms of the four ABJM charges $\Delta_I$, the superpotential term fixes one charge to unity, so we describe it~by 
\begin{align}
\label{eq:slice}
  (\Delta_1,\Delta_2,\Delta_3,\Delta_4) &= (1,\delta_1,\delta_2,\delta_3) \, , \nn \\ 
  \delta_1+\delta_2+\delta_3 &= 1 \, , \quad \delta_i>0 \, .
\end{align}
Note that the Nosaka locus \eqref{eq:NosakaLocus} intersects this only at its boundary.
In the following we restrict to $k=1$ and the round sphere. 

\paragraph{Mirror Fermi gas.}
We use $3d$ mirror symmetry to relate ABJM with $k=1$ to the ADHM theory with $N_f = 1$, with charges $\widetilde{\boldsymbol{\Delta}} = (1, \delta_2, \delta_1, \delta_3)$ obtained using the parameter map in (2.45) of \cite{Bobev:2025ltz}.
This also means that the R-charges of the three adjoint chiral multiplets are $\boldsymbol{\Delta_a} = (1, \delta_2, 1-\delta_2)$, the R-charge of the fundamental (and anti-fundamental) is $r = (1+\delta_2)/2$, and the monopole is $\Delta_m = (\delta_3-\delta_1)/2$.
We then specialize the matrix model integral in (2.42) of \cite{Bobev:2025ltz} and use double sine function identities to rewrite it as
\begin{align}
\label{eq:ZADHM_AlmostGas}
	Z^{\rm ADHM} &= \frac{1}{N!} \int \prod_{i=1}^N \dd x_i  \,  W(x_i) 
	\frac{ \prod_{i>j}4\sinh^2(\pi x_{ij}) }{ \prod_{i,j} 2 \cosh \left[ \pi \left( x_{ij} + \ii \left(  \frac{1}{2} - \delta_2 \right) \right) \right]}  \, , \nn \\[5pt]
	W(x_i) &\equiv \ex^{ - 2\pi \Delta_m x_i } \frac{ s_1 \left( x_i + \ii \frac{1-\delta_2}{2} \right) }{ s_1 \left( x_i - \ii \frac{1-\delta_2}{2} \right) }  \, ,
\end{align}
where $x_i \equiv \mu_i / (2\pi)$ are real.

The integrand can be rewritten as the determinant of a one-particle kernel, and thus admits an interpretation as a Fermi gas, as noted in \cite{Marino:2015ixa}.
In particular, we can write
\begin{equation}
	W(x_i) = \lvert \Psi_{\delta_1/2, \delta_3/2}(x_i) \rvert^2 \, ,
\end{equation}
where
\begin{equation}
\label{eq:DefPsiKM}
	\Psi_{a,c}(x) \equiv \frac{\ex^{2\pi ax}}{ \Phi_1 ( x - \ii (a+c) ) } \, , 
\end{equation}
is a nowhere-vanishing function of real $x$, the parameters $a$ and $c$ are non-negative real numbers such that $a+c<1$, and $\Phi_1(x)$ is Faddeev's quantum dilogarithm $\Phi_b(x)$ specialized to $b=1$.  In our case, $a=\delta_1/2$ and $c=\delta_3/2$.
This is a crucial step to show that $Z^{\rm ADHM}$ in \eqref{eq:ZADHM_AlmostGas} is in fact
\begin{align}
\label{eq:ZADHM_Gas}
	&Z^{\rm ADHM} \Big( N; \widetilde{\boldsymbol\Delta} = ( 1 , \delta_2 ,\delta_1 , \delta_3 ) \Big) = 
	\frac{1}{N!} \int \dd^Nx \, \det_{i,j} \left[ \rho_{\delta_1/\delta_3,\delta_2/\delta_3}(x_i,x_j) \right]  \, ,
\end{align} 
where the kernel is the operator
\begin{equation}
\label{eq:KMkernel}
	\rho_{m,n}(x,y) = \frac{\overline{\Psi_{a,c}(x)} \Psi_{a,c}(y)} {2 \cosh \left[ \pi (x-y+ \ii(a+c-nc)) \right]} \, , 
\end{equation}
introduced and studied in \cite{Kashaev:2015kha}.
This operator is the inverse of $\mathsf{O}_{m,n}(\hbar) = \ex^{\mathsf{x}} + \ex^{\mathsf{y}} + \ex^{-m \mathsf{x} - n \mathsf{y}}$ with $[\mathsf{x}, \mathsf{y}] = \ii \hbar$ and $m,n\in\R_{>0}$. If we specialize further $m,n \in \Z_{>0}$, the latter operator has a geometric interpretation as the quantization of the mirror curve of the local weighted projective space $\mc{O}(-1-m-n) \to \mathbb{CP}(1,m,n)$.  In our case, 
\begin{align}\label{mndvars}
m= \frac{\delta_1}{\delta_3}\,,\qquad n=\frac{\delta_2}{\delta_3}\,,\qquad \hbar = 2\pi \delta_3\,.
\end{align}

There is a special value $m=n=1$, corresponding to
$\delta_1=\delta_2=\delta_3=\frac13$, where there is an additional $\mathbb Z_3$
symmetry and $\mathsf O_{1,1}$ is the quantization of the mirror curve
of local $\mathbb{CP}^2$. This is precisely the superconformal point of
mABJM, for which $\hbar=2\pi/3$.

\paragraph{Airy coefficients.}
Importantly, the perturbative grand potential of the spectral problem of $\rho_{m,n}$ has been studied in \cite{Hatsuda:2015oaa}, and it was found to have the standard Airy form \eqref{Jpert}, with
\begin{align}
\label{eq:Hatsuda_ABC}
	\cC &= \frac{(m+n+1)^2}{4\pi mn\hbar} \, , \nn \\
	\cB &= \frac{ \pi (m^2+mn+n^2+m+n+1)}{12mn\hbar} - \frac{(m+n+1)\hbar}{48\pi } \,, \nn \\
	\cA &= \frac{1}{4} \Big[ A\Big( \frac{\hbar}{\pi} \Big) + A\Big( m\frac{\hbar}{\pi} \Big) + A\Big( n\frac{\hbar}{\pi} \Big) 
	- A\Big( (m+n+1)\frac{\hbar}{\pi} \Big) \Big] \, ,
\end{align}
where the last expression had been conjectured and checked via WKB until order $\hbar^5$, and $A$ is the contribution of the constant map in topological strings given in \eqref{eq:ABJMconstantmap}.
Substituting the values \eqref{mndvars} gives
\begin{align}
\label{eq:Hatsuda_ABC_Us}
	\cC &= \frac{1}{8\pi^2 \delta_1\delta_2\delta_3} \, , \nn \\
	\cB &= \frac{1+\delta_1^2+\delta_2^2+\delta_3^2}{48 \delta_1\delta_2\delta_3} - \frac{1}{24} \,, \nn \\
	\cA &= \frac{1}{4} \left[ A\left( 2\delta_3 \right) + A\left( 2\delta_1 \right) + A\left( 2\delta_2 \right) - A\left( 2 \right) \right] \, .
\end{align}
We stress that these values have been obtained from the spectral problem for the Fermi gas equivalent to ADHM. When comparing with mABJM, we should in principle be mindful of potential phase differences in the partition functions of the mirror theories, due to decoupled $U(1)$ factors.

In fact, these do not appear with the ABJM--ADHM mirror normalization we choose, though in general matching real Airy coefficients does not independently rule out a possible phase.
First, the values for $\cC$ and $\cB$ in \eqref{eq:Hatsuda_ABC_Us} match those obtained using large $N$ methods for mABJM in \cite{Bobev:2025ltz}, after specialization to $b=1$, $k=1$, and \eqref{eq:slice}.
After the same specialization, these values also match those obtained from equivariant localization in M-theory in \eqref{eq:ABJMCB}.
Moreover, the character \eqref{eq:ABJMk1decay} takes the form
\begin{align}
	\mc{K}(t) &= \frac{\sum_{i=1}^3 \coth(\delta_i t/2)-\coth(t/2)}{8\sinh^2(t/2)} \nn \\
	&= \frac{1}{4} \left[ \mc{K}_{2\delta_1}(t) + \mc{K}_{2\delta_2}(t) +\mc{K}_{2\delta_3}(t) - \mc{K}_2(t) \right] \, ,
\end{align}
where $\mc{K}_\kappa(t)$ is the round ABJM character \eqref{eq:ABJMroundBarnes} analytically continued to positive real $\kappa$.
The Mellin integrals share a convergence domain ${\operatorname{Re}} \, s > 3$, so linearity followed by meromorphic continuation shows that we find the same value of $\cA$ obtained by Hatsuda in \eqref{eq:Hatsuda_ABC_Us}.

\section{Type IIA reduction of the ABJM one-loop character}
\label{sec:IIAreduction}

In this appendix we identify more explicitly the geometric origin of the Type IIA
expansion described in the main text. The basic idea is to promote the
$\mathbb Z_k$ action to its parent M-theory circle $U(1)_{\rm M}$ and
rewrite the Molien sum as a contour integral. The resulting residues perform
the equivariant push-forward along this circle: one complex direction is
removed from both the tangent denominator and the $T-T^*$ numerator,
leaving the tangent-cotangent character of the K\"ahler Calabi--Yau
three-fold quotient (the resolved conifold). This allows us to derive an explicit expression for the
equivariant characteristic number $\mathcal E_g^{\rm red}$ appearing in
\eqref{eq:IIAHodge}.

\paragraph{Circle reduction of the character.}
Recall that the $\mathbb Z_k$ action on the four internal complex directions is
the subgroup of $U(1)_{\rm M}$ with charges
\begin{align}
(c_1,c_2,c_3,c_4)
&=
(1,1,-1,-1)\,.
\label{eq:IIAcharges}
\end{align}
The corresponding additive weights are
$u_I=-Q\Delta_I/2$ as in \eqref{eq:ABJMnormalizedweights}, so that the multiplicative
weights are $q_I=\ex^{Q\Delta_I t/2}$. Introducing a fugacity $z$ for
$U(1)_{\rm M}$ amounts to replacing $q_I\to z^{c_I}q_I$. If $f(z)$
denotes the resulting character, the Molien average in
\eqref{eq:ABJMMolien} may be written as
\begin{align}
\frac1k\sum_{z^k=1}f(z)
&=
\sum_{z^k=1}
\mathop{\rm Res}_{z}
\bigg[
\frac{\dd z}{2z}
\frac{z^k+1}{z^k-1}
f(z)
\bigg] .
\label{eq:IIAMoliencontour}
\end{align}
Indeed, one easily checks that the residue of the kernel at each $k$th
root of unity is $1/k$. For the charge assignment
\eqref{eq:IIAcharges} there are no residues at $z=0,\infty$, so the
global residue theorem gives
\begin{align}
\frac1k\sum_{z^k=1}f(z)
&=
-\sum_{I=1}^4
\mathop{\rm Res}_{z=z_I}
\bigg[
\frac{\dd z}{2z}
\frac{z^k+1}{z^k-1}
f(z)
\bigg] ,
\label{eq:IIAinternalpoles}
\end{align}
where $z_I^{c_I}q_I=1$ and hence
$z_I=\exp\left(-\frac{Q\Delta_I t}{2c_I}\right)$. At the $I$th pole the corresponding internal
tangent denominator is removed, while crucially the contribution of this
direction to the tangent-cotangent numerator vanishes,
\begin{align}
(z_I^{c_I}q_I)^{-1}-z_I^{c_I}q_I=0\,.
\label{eq:IIAverticalnumerator}
\end{align}
Thus the $U(1)_{\rm M}$ direction is removed simultaneously from the
denominator and from the $T-T^*$ numerator. A short computation gives
the remaining factor from the contour kernel,
\begin{align}
\mathop{\rm Res}_{z=z_I}
\bigg[
\frac{\dd z}{2z}
\frac{z^k+1}{z^k-1}
\frac{1}{1-z^{c_I}q_I}
\bigg]
&=
\frac12\coth\bigg(\frac{k Q\Delta_I t}{4}\bigg)\, .
\label{eq:IIAkernelresidue}
\end{align}

Having effectively projected out the M-theory circle, for $j\neq I$ the
remaining quotient weights are
\begin{align}
\widetilde u_{j|I}
&=
u_j-\frac{c_j}{c_I}u_I\,.
\label{eq:IIAquotientweights}
\end{align}
For \eqref{eq:IIAcharges} these are $(-d_I,-r_I,-s_I)$, with
$d_I,r_I,s_I$ defined in \eqref{eq:IIAweights}. Together with the two
external weights, the reduced tangent weights at the $I$th residue are
hence
\begin{align}
\label{eq:app_ReducedWeights}
\left(b,b^{-1},-d_I,-r_I,-s_I\right),
\end{align}
with $d_I+r_I+s_I=Q$. The corresponding tangent-cotangent character is
$-h_I(t)$, with $h_I(t)$ given in \eqref{eq:IIAhI}. Combining this minus
sign with that in the global residue theorem
\eqref{eq:IIAinternalpoles}, and using \eqref{eq:IIAkernelresidue},
gives precisely \eqref{eq:IIAcharacter}. Thus the exact finite $k$
decomposition has a direct interpretation as equivariant reduction along
the M-theory circle.\footnote{Relatedly, the local reduced character $h_I(t)$ can also be directly obtained from the complex of twisted supergravity on $\C^5\times S^1$. The single particle index found in \cite{Raghavendran:2021qbh} has the form
$i_{\rm SP} = \frac{\sum_{i=1}^5 \left( q_i - q_i^{-1} \right) }{\prod_{i=1}^5(1-q_i)}$
and substituting $q_i=\ex^{-tw_i}$ with $w_i$ from \eqref{eq:app_ReducedWeights} gives exactly $h_I(t)$ in \eqref{eq:IIAhI}.
Indeed, this has the form of the index \eqref{eq:character-general} on $\C^5$, that is, $i_{\rm SP} = - \cK_{\C^5}$, with the additional $SU(5)$ constraint $\prod_i q_i = 1$ coming from the twist. However, this is not a straightforward reduction from $\cK_{\C^{6}}(t)$ by simply setting $q_6=1$. Instead, once the eleven-dimensional theory is represented as the boundary of a twelve-dimensional relative problem, the index-theoretic avatar of the circle reduction is the push-forward along the disk that fills the circle.}

\paragraph{Reduced geometry and RR flux.}
We next describe the quotient geometry entering this reduction. With the
normalization
\begin{align}
\mu_{\rm M}
&=
|z_1|^2+|z_2|^2-|z_3|^2-|z_4|^2\,,
\end{align}
the K\"ahler quotient at level $\zeta$ is
\begin{align}
\Cthree_\zeta
&\equiv
\C^4//_\zeta\mskip1mu U(1)_{\rm M}
=
\mu_{\rm M}^{-1}(\zeta)/U(1)_{\rm M}\,.
\label{eq:IIAKahlerquotient}
\end{align}
At $\zeta=0$, $\Cthree_0$ is the singular conifold. For either sign
$\zeta\neq0$, $\Cthree_\zeta$ is the resolved conifold
\begin{align}
\Cthree_\zeta
&\cong
\mathcal O(-1)\oplus\mathcal O(-1)
\longrightarrow \mathbb{CP}^1\,,
\end{align}
with the two signs of $\zeta$ giving the two resolutions related by the
flop. Equivalently, the real quotient $\C^4/U(1)_{\rm M}$ is fibred over
the moment map coordinate $\zeta\in\R$, with fibre $\Cthree_\zeta$ and
K\"ahler parameter varying linearly with $\zeta$
\cite{Martelli:2009ga}.
Each resolved fibre has two toric fixed points, namely the two poles of
the zero-section $\mathbb{CP}^1$. We denote these by $p_I$. In the chamber
$\zeta>0$ they are $p_1,p_2$, while for $\zeta<0$ they are $p_3,p_4$.
These are naturally identified with the four residues above: the residue
labelled by $I$ is the fixed point contribution from $p_I$ in the
corresponding K\"ahler chamber.

Let $\LM$ denote the primitive line bundle associated with the
$U(1)_{\rm M}$ circle. On either resolved fibre it has unit first Chern
class through the zero-section,
\begin{align}
\int_{\mathbb{CP}^1}c_1(\LM)&=1\,,
\end{align}
with the orientation chosen accordingly. After the
$\mathbb Z_k\subset U(1)_{\rm M}$ quotient, the physical M-theory circle
bundle is
\begin{align}
\LM^{(k)}
&\cong
\LM^{\otimes k}\,,
\end{align}
and hence, in conventions where the constant normalization $g_s$ is set
to one,
\begin{align}
c_1\big(\LM^{(k)}\big)
&=
\left[\frac{F_2}{2\pi\ell_s}\right]
=
k\,c_1(\LM)\,,
\label{eq:IIAflux}
\end{align}
where $\ell_s$ is the string length. Thus the RR two-form $F_2$ carries
$k$ units of flux through the exceptional $\mathbb{CP}^1$.

At the fixed point $p_I$, the equivariant weight of the primitive circle
bundle is
\begin{align}
c_1^T(\LM)\big|_{p_I}
&=
-\frac{u_I}{c_I}
=
\frac{c_IQ\Delta_I}{2}\,.
\end{align}
The three tangent weights of $T\Cthree_\zeta$ at the same fixed point are
$(-d_I,-r_I,-s_I)$, and hence
\begin{align}
c_1^T(T\Cthree_\zeta)\big|_{p_I}
&=
-d_I-r_I-s_I=-Q\,.
\end{align}
It is therefore useful to define
\begin{align}
\rflux
&\equiv
-\frac{2}{k}\,
\frac{c_1^T\big(\LM^{(k)}\big)}
     {c_1^T(T\Cthree_\zeta)}
=
-2\,
\frac{c_1^T(\LM)}
     {c_1^T(T\Cthree_\zeta)}\,,
\label{eq:IIAfluxclass}
\end{align}
so that
\begin{align}
\rflux\big|_{p_I}
&=
c_I\Delta_I\,.
\label{eq:IIAfluxfixedpoint}
\end{align}
All ratios and negative powers of equivariant classes below are understood
in localized equivariant cohomology. The sign in
\eqref{eq:IIAfluxfixedpoint} distinguishes the two K\"ahler chambers, but
drops out of the power $\rflux^{\,2-2g}$ appearing below.

\paragraph{Equivariant constant maps.}
We now derive the form of \eqref{eq:IIAHodge} directly from the reduced
character. To isolate the $n=0$ term in \eqref{eq:IIAgrading}, write
$d_I=Q\delta_I$, $r_I=Q\rho_I$, $s_I=Q\sigma_I$, with
$\delta_I+\rho_I+\sigma_I=1$, and formally expand $h_I(t)$ in $Q$ at
fixed $(\delta_I,\rho_I,\sigma_I)$. This is simply an algebraic extraction
of the finite refinement grading in \eqref{eq:IIAgrading}. The $Q^0$
term is
\begin{align}
\left.h_I(t)\right|_{Q^0}
&=
\frac{(d_I+r_I)(d_I+s_I)(r_I+s_I)}
     {d_Ir_Is_I}
\bigg[
\frac{1}{4\sin^2(t/2)}
-\frac{1}{12}
\bigg]
-\frac{1}{12}\,.
\label{eq:IIAhQzero}
\end{align}

The three equivariant Chern roots of $T\Cthree_\zeta$ at $p_I$ are simply the weights
$(-d_I,-r_I,-s_I)$. For a complex rank three bundle one has
\begin{align}
\Lambda^2T\Cthree_\zeta
&\cong
T^*\Cthree_\zeta\otimes\det T\Cthree_\zeta\,,
\label{eq:IIArankthree}
\end{align}
so the Chern roots of $\Lambda^2T\Cthree_\zeta$ are their three pairwise
sums. Hence
\begin{align}
\frac{
c_3^T\!\left(\Lambda^2T\Cthree_\zeta\right)|_{p_I}
}{
e_T(T\Cthree_\zeta)|_{p_I}
}
&=
\frac{(d_I+r_I)(d_I+s_I)(r_I+s_I)}
     {d_Ir_Is_I}\,.
\label{eq:IIALambda2ratio}
\end{align}
Equivalently,
$c_3^T(\Lambda^2T\Cthree_\zeta)
=c_1^Tc_2^T-c_3^T$, where the Chern classes on the right-hand side are
those of $T\Cthree_\zeta$. Notice that although $\Cthree_\zeta$ is
Calabi--Yau, $\det T\Cthree_\zeta$ carries non-zero equivariant weight,
$c_1^T(T\Cthree_\zeta)|_{p_I}=-Q$. 

Using
\begin{align}
\frac{1}{4\sin^2(t/2)}-\frac{1}{12}
&=
\frac{1}{t^2}
+
\sum_{g\geq2}
\frac{|B_{2g}|}{2g(2g-2)!}\,t^{2g-2}\,,
\end{align}
we therefore obtain, for $g\geq2$,
\begin{align}
\left.h_{I,2g-2}\right|_{Q^0}
&=
\frac{|B_{2g}|}{2g(2g-2)!}\,
\frac{
c_3^T\!\left(\Lambda^2T\Cthree_\zeta\right)|_{p_I}
}{
e_T(T\Cthree_\zeta)|_{p_I}
}\,.
\label{eq:IIAhgeometric}
\end{align}
Substituting this into \eqref{eq:IIAAg}
gives
\begin{align}
\cU_{g,0}
&=
\frac{2^{2g-2}\zeta(2g-2)|B_{2g}|}
     {2g(2g-2)}\,
\mathcal E_g^{\rm red}\,,
\label{eq:IIAUg0}
\\
\mathcal E_g^{\rm red}
&\equiv
\sum_{I=1}^4
\left.\rflux^{\,2-2g}\right|_{p_I}
\frac{
c_3^T\!\left(\Lambda^2T\Cthree_\zeta\right)|_{p_I}
}{
e_T(T\Cthree_\zeta)|_{p_I}
}\,.
\label{eq:IIAEredFP}
\end{align}
Here we used \eqref{eq:IIAfluxfixedpoint}; since $2-2g$ is even, the
sign distinguishing the two K\"ahler chambers drops out.

Writing $\Cthree_\pm$ for the two resolved conifold chambers,
\eqref{eq:IIAEredFP} is equivalently the localized equivariant integral
\begin{align}
\mathcal E_g^{\rm red}
&=
\sum_{\epsilon=\pm}
\int_{\Cthree_\epsilon}^{T}
\rflux^{\,2-2g}\,
c_3^T\!\left(\Lambda^2T\Cthree_\epsilon\right)\,.
\label{eq:IIAgeometric}
\end{align}
Using the standard formula for $\zeta(2g-2)$, equations
\eqref{eq:IIAFred} and \eqref{eq:IIAUg0} then give
\begin{align}
\widehat F^{\rm red}_{g,0}
&=
\kappa_g
\sum_{\epsilon=\pm}
\int_{\Cthree_\epsilon}^{T}
\rflux^{\,2-2g}\,
c_3^T\!\left(\Lambda^2T\Cthree_\epsilon\right)\,,
\quad g\geq2\,,
\label{eq:IIAfluxconstantmap}
\end{align}
with $\kappa_g$ precisely the Hodge integral coefficient in
\eqref{eq:IIAHodge}.

Equation \eqref{eq:IIAfluxconstantmap} therefore has exactly the structure
of the usual degree-zero topological string result: a universal
worldsheet Hodge integral multiplying a target space characteristic
number. Here this becomes an equivariant invariant dressed by the normalized RR-flux class $\rflux$ through the factor $\rflux^{\,2-2g}$. In particular, the
class $c_3^T(\Lambda^2T\Cthree)$ follows directly from reduction of the
eleven-dimensional $T-T^*$ complex. This differs from the equivariant
constant map extensions proposed in
\cite{Cassia:2025aus,Cassia:2025jkr}, where the corresponding
higher-genus target space class is the equivariant Euler class:
the circle reduction here instead fixes the combination
$\rflux^{\,2-2g}\,
c_3^T(\Lambda^2T\Cthree)$.

The same reduced character also determines the higher-refinement components $U_{g,n}$, $n>0$, by retaining the corresponding higher powers of $Q^2$ in the expansion leading to \eqref{eq:IIAhQzero}. Thus the circle reduction captures the complete finite refinement tower in \eqref{eq:IIAgrading}, while the $n=0$ component is distinguished by the simple characteristic class and Hodge integral interpretation above.

Our result strongly suggests that there should be an intrinsic formulation of
the same protected sector directly in Type IIA string theory on the
resolved conifold in the presence of RR two-form flux. The
M-theory calculation above gives a precise prediction for the resulting
equivariant, flux-dependent constant map invariants, including their
dependence on the two flop chambers.

The same construction should apply also to the
$V^{5,2}/\mathbb Z_k$ example. In this case the K\"ahler quotient of the
Calabi--Yau four-fold by $U(1)_{\rm M}$ gives the three-fold
$W_2^\zeta$ of \cite{Martelli:2009ga}: the singular fibre $W_2^0$ is
the Laufer three-fold singularity, while $\zeta\neq0$ gives its two
small resolutions. Thus the $V^{5,2}/\Z_k$ one-loop character should
similarly determine an equivariant, RR-flux-dressed topological string
invariant on the corresponding resolved three-folds.

\section{Unrefined topologically twisted index}
\label{sec:TTIunrefined}

The ordinary TTI is the $\epsilon\to0$ limit of the refined TTI.
In this appendix we derive \eqref{eq:TTIallorders} from
\eqref{eq:TTIcanonical} and obtain a closed form expression for the
$N$-independent constant
$\widehat f_0(k,\bm\Delta,\bm{\mf{n}})$ for generic chemical potentials
and magnetic charges.

The block integrals in \eqref{eq:TTIcanonical} are Airy-type integrals and are exact at finite $\epsilon$. To extract their unrefined limit, we use their asymptotic expansion as $\epsilon\to0$.
For each $\sigma=\pm$, we take
\begin{align}
\mu_\sigma^*
=
\sigma\sqrt{\frac{N-\cB_\sigma}{\cC_\sigma}}\,,
\end{align}
to be the saddle about which we expand.\footnote{These saddle choices are those associated with the contour prescription used in defining the block integrals. An \textit{a priori} derivation of this prescription remains to be established.}
Since the exponent is cubic, its expansion about this saddle is exact:
\begin{align}\label{saddlept}
 -I^\sigma_{\rm SUSY}(\mu_\sigma)-N\mu_\sigma
 &=
 -\frac{2}{3}\cC_\sigma(\mu_\sigma^*)^3
 +\cC_\sigma\mu_\sigma^*
   (\mu_\sigma-\mu_\sigma^*)^2
 +\frac{\cC_\sigma}{3}
   (\mu_\sigma-\mu_\sigma^*)^3 \,.
\end{align}
Rescaling via $t_\sigma=(\cC_\sigma\mu_\sigma^*)^{1/2}(\mu_\sigma-\mu_\sigma^*)$
the last two terms become  
$t_\sigma^2+ \lambda_\sigma{t_\sigma^3}$ where 
$\lambda_\sigma=\frac{1}{3\sqrt{\cC_\sigma(\mu_\sigma^*)^3}}$.
Since $\cC_\sigma=O(\epsilon^2)$ and $\mu_\sigma^*=O(\epsilon^{-1})$, $\lambda_\sigma$ is $O(\epsilon^{1/2})$.
The term linear in $\lambda_\sigma$ is odd in $t_\sigma$ and hence integrates to zero, so the first non-vanishing correction to the Gaussian result is $O(\epsilon)$.
It follows that the logarithm of the partition function has the
asymptotic form 
\begin{equation}\label{TTImid}
 \log Z_{\rm M}^{\rm pert}[N]
 =
 \sum_\sigma
 \left[
 -\frac{2}{3}\cC_\sigma(\mu_\sigma^*)^3
 -\frac{1}{2}\log(4\pi\cC_\sigma\mu_\sigma^*)
 +\cA_\sigma
 \right]
 +O(\epsilon)\, .
\end{equation}
Thus \eqref{TTImid} gives the exact $\epsilon\to0$ limit of the Airy block integrals; at finite $\epsilon$ the omitted terms are the higher terms in their Airy asymptotic expansion. The first term in \eqref{TTImid} gives precisely the first line of \eqref{eq:TTIallorders}, while the Gaussian contribution is
\begin{align}\label{TTIgaussian}
	& - \frac{1}{2} \sum_\sigma \log(4\pi \cC_\sigma \mu_\sigma^*) = - \frac{1}{2} \log \widehat{N}_{k,\Delta} 
	- \frac{1}{2} \log \frac{32}{k \prod_I \Delta_I} - \frac{1}{2} ( \log\epsilon + \log(-\epsilon)) +O(\epsilon^2)\,.
\end{align}

We now turn to $\cA_\sigma$. To expand \eqref{eq:TTIBarnes} in $\epsilon$, we introduce the five-fold Barnes zeta 
\begin{equation}
 \zeta_5^{(r)}(s,z)
 \equiv
 \sum_{m_1,m_3,\dots,m_6}
 \frac{
 \omega^{-r\sum_{I=1}^{4}c_{I+2}m_{I+2}}
 }
 {(z+\bm m\cdot\bm\Omega_5)^s}\, ,
 \label{eq:twistedBarnes5}
\end{equation}
 for $z\neq0$, where $\bm\Omega_5 \equiv (1, \Delta_1/2,\dots,\Delta_4/2)$.
For $z=0$, we use the same definition with the zero lattice vector omitted.
The six-fold Barnes zeta \eqref{eq:twistedBarnes} with periods \eqref{eq:TTIOmega} can then be expanded as
\begin{align}
	& \zeta_6^{(r)}(s,z+\epsilon) = \frac{\zeta_5^{(r)}(s-1,z)}{\epsilon (s-1)} - \frac{1}{2} \zeta_5^{(r)} (s,z) 
	+ \sum_{I=1}^4 \mf{n}_I \partial_{\Delta_I}  \frac{\zeta_5^{(r)}(s-1,z)}{s-1} + O(\epsilon) \,.
\end{align}
For $z=0$, there is an additional contribution to the expansion from the 
mode $m_1 = m_3 = \dots = m_6 = 0$ in the six-fold Barnes zeta,
given by $\sum_{m_2} [\epsilon (m_2+1)]^{-s} = \epsilon^{-s} \zeta(s)$.

It is useful to organize the expansion of $\cA_+$ in the following form,
\begin{align}
	\cA_+ & = -\frac{\widehat{g}_0(k,\bm\Delta)}{\epsilon} + \frac{1}{2} \log \epsilon - \frac{1}{2}\log(2\pi) - \sum_I \mf{n}_I \partial_{\Delta_I} \widehat{g}_0 \nn\\
	&\quad  \ + \frac{1}{k} \partial_s \left. \sum_{r=0}^{k-1} \Big[ 2 \zeta_5^{(r)}(s,1) - \frac{1}{2}v_r(s) \Big] \right\vert_{s=0} + O(\epsilon) \,,
\end{align}
where we have included the contribution from the zero-lattice mode $\partial_s(\epsilon^{-s} \zeta(s))|_{s=0} = \frac{1}{2} \log \epsilon - \frac{1}{2} \log (2\pi)$.
From expanding \eqref{eq:TTIBarnes}, we find
\begin{align}\label{g0general}
	\widehat{g}_0(k,\bm\Delta) & = - \left. \frac{1}{k} \partial_s \sum_{r=0}^{k-1} \frac{v_r(s-1)}{s-1} \right\vert_{s=0} \,, \nn \\
	v_r(s) & \equiv \zeta_5^{(r)}(s,0) - \zeta_5^{(r)}(s,2) \nonumber\\
&\qquad + \sum_{I=1}^4 \omega^{-r c_{I+2}} \zeta_5^{(r)}\!\Big(
s,\Big(1+\frac{\Delta_I}{2}\Big)\Big)
-\sum_{I=1}^{4}\omega^{r c_{I+2}}
\zeta_5^{(r)}\!\Big(
s,\Big(1-\frac{\Delta_I}{2}\Big)\Big) \,.
\end{align}
The second block $\cA_-$ is obtained from $\cA_+$ by $\epsilon \mapsto - \epsilon$.
Combining the above as in \eqref{TTImid}, we use the same
branches of $\log\epsilon$ and $\log(-\epsilon)$ in the
Gaussian and one-loop contributions. These logarithmic
terms therefore cancel, giving the following closed form
expression for the $N$-independent constant term in \eqref{eq:TTIallorders}:
\begin{align}\label{f0general}
	\widehat{f}_0(k,\bm\Delta,\bm{\mf{n}}) & = -2 \sum_I \mf{n}_I \partial_{\Delta_I} \widehat{g}_0(k,\bm\Delta) + \frac{1}{2} \log \frac{k \prod_I \Delta_I}{128 \pi^2} 
	\nn \\
	&\qquad 
	+ \frac{2}{k} \partial_s \left. \sum_{r=0}^{k-1} \Big[ 2 \zeta_5^{(r)}(s,1) - \frac{1}{2} v_r(s) \Big] \right\vert_{s=0}  \,.
\end{align}

The appearance of $\widehat g_0(k,\bm\Delta)$ as the coefficient of the leading $1/\epsilon$ 
term has a natural field theory interpretation.
In the
$\epsilon\to0$ limit, the leading $1/\epsilon$ term of the refined
topologically twisted blocks is governed by the Bethe potential.
Accordingly, the same quantity $\widehat g_0(k,\bm\Delta)$ appears as
the $N$-independent perturbative term in its large $N$ expansion
\cite{Bobev:2022wem},
\begin{align}\label{Bethe}
\frac{1}{2\pi}\operatorname{Im}\mathcal V(N,k,\bm\Delta)
&=
\frac{\pi\sqrt{2k\prod_I\Delta_I}}{3}
\widehat N_{k,\Delta}^{3/2}
+\widehat g_0(k,\bm\Delta)
+\widehat g_{\rm np}(N,k,\bm\Delta)\,.
\end{align}
Here $\widehat g_{\rm np}$ denotes non-perturbative corrections.

\paragraph{Field theory checks.}
We can compare the general expressions \eqref{g0general} and
\eqref{f0general} with both analytic and numerical field theory results. 
For $\widehat{g}_0$, setting $\Delta_I = \frac{1}{2}$ in \eqref{g0general} and performing the root-of-unity sum,
we may write $v_r(s)$ at the superconformal point as
\begin{equation}
	\frac{1}{k} \sum_{r=0}^{k-1} v_r^{\rm sc}(s) = \zeta_{V_k}(s) \,,
\end{equation}
where
\begin{align}
	V_k(t) & \equiv \coth(kt/8) (2 \coth(t/4) - \coth(t/2)) 
	+ \frac{k}{2\sinh^2(kt/8)} - 1 \,.
\end{align}

The Mellin transform can be simplified using integration by parts:
\begin{align}
	\widehat{g}_0(k,\bm\Delta^{\rm sc}) & = \partial_s \left. \left[ - \frac{1}{\Gamma(s)} \int_0^\infty \rd t \ t^{s-2} V_k(t) \right] \right\vert_{s=0} \nn \\
	& = \partial_s \left. \left[ - \frac{1}{(s-1) \Gamma(s)} \left( \left. t^{s-1} V_k(t) \right\vert_0^\infty - \int_0^\infty \rd t \ t^{s-1} V_k'(t) \right) \right] \right\vert_{s=0} \nn \\
	& = -\left.  \partial_s \left[ \frac{1}{\Gamma(s)} \int_0^\infty \rd t \ t^{s-1} V_k'(t) \right]\right\vert_{s=0} \,.
\end{align}
The boundary term vanishes for $\operatorname{Re}s>3$,
and the result is then continued meromorphically to $s=0$.\footnote{The boundary term vanishes for $\operatorname{Re} s>3$, since as $t \rightarrow \infty$, $V_k(t)$ goes to zero, and as $t \rightarrow 0$, $V_k(t) \sim t^{-2}$ so $t^{s-1} V_k(t) \sim t^{s-3}$.}
To arrive at the last line we made use of the fact that $V_k'(-t) = - V_k'(t)$, hence $\zeta_{V_k'}(0) = 0$. 
In terms of \eqref{eq:ABJMroundBarnes}, $V_k'(t)$ can be written as
\begin{align}
	V_k'(t) &=- \cK_k(t/2) + \cK_{k/2}(t) - \frac{k}{2} \cK_{4/k}(kt/4) 
	+ \frac{k}{4} \cK_{8/k}(k t/4) - \frac{k^2}{4} \cK_2 (kt/4) \,.
\end{align}
It follows that
\begin{align}
	\widehat{g}_0(k,\bm\Delta^{\rm sc}) & = A(k) - A(k/2) + \frac{k}{2} A(4/k)
	- \frac{k}{4} A(8/k) + \frac{k^2}{4} A(2) \,,
\end{align}
in agreement with equation (32) of \cite{Hong:2026zul}, where $A(2) = - \frac{\zeta(3)}{2\pi^2}$ (see also \cite{Hosseini:2026dkj}).
For generic flavour chemical potentials, we have checked numerically\footnote{For these numerical checks we use the finite-part implementation of the Mellin continuation described in footnote~\ref{footfive}.}
that our general expression for
$\widehat g_0$ agrees with the numerical values extracted from the ABJM
Bethe potential in appendix C.2 of~\cite{Bobev:2022wem}, in the course of
their analysis of the Cardy-like superconformal index.

Similarly, for $\widehat{f}_0$, \eqref{f0general} at the superconformal point $\Delta^{\rm sc}_I=\mathfrak n^{\rm sc}_I=\tfrac12$
can be expressed in terms of the character
\begin{align}
	\cK_{f}(t) & = - 4 \cK_k(t/2) - 3k \cK_{4/k} (kt/4) + \frac{3k}{2} \cK_{8/k}(kt/4)
	- \frac{k^2}{2} \cK_2(kt/4) - \frac{2}{\ex^{kt/4}-1} \,,
\end{align}
such that
\begin{equation}
	 \widehat f_0(k,\bm\Delta^{\rm sc},\bm{\mf{n}}^{\rm sc}) = \zeta'_{\cK_f}(0) + \frac{1}{2} \log\frac{k}{2048\pi^2} \,.
\end{equation}
Using the ABJM constant map function $A(k)$, this gives precisely
\eqref{eq:TTIf0}, in agreement with equation (42) of
\cite{Hong:2026zul} (see also \cite{Hosseini:2026lvj}).

For general flavour chemical potentials and magnetic charges, our expression
for $\widehat f_0$ agrees with the numerical results in appendix C
of~\cite{Bobev:2022eus}, after accounting for the convention
$\widehat f_0^{\rm there}=\widehat f_0^{\rm here}+\log k$.
The additional $\log k$ in~\cite{Bobev:2022eus} accounts for the
$k$-fold degeneracy of the Bethe vacua, which is not included in the
normalization used in~\eqref{eq:TTIcanonical}.

\section{The \texorpdfstring{$V^{5,2}$}{V52} superconformal specialization}
\label{sec:V52appendix}

We first clarify the role of resolutions in the construction of the
$V^{5,2}$ one-loop character. On a proper equivariant resolution
$\pi:\widetilde X\to X$, one localizes the pull-back of the intrinsic
virtual class
$T_X^{\rm vir}-(T_X^{\rm vir})^*$, rather than
$T\widetilde X-T^*\widetilde X$. 
For cones with rational singularities, the resulting equivariant index is
independent of the choice of resolution; this follows from
$R\pi_*\mathcal O_{\widetilde X}\simeq\mathcal O_X$
together with the projection formula.
In the present case the blow-up
of the $V^{5,2}$ cone is non-crepant, so replacing the pulled-back
virtual class by $T\widetilde X-T^*\widetilde X$ would instead generate
an additional contribution from the exceptional divisor.

At the superconformal point $m_1=m_2=0$, \eqref{eq:V52CB} gives
$\cB=-5/12+3/Q^2$ and $\cC=81/(4\pi^2Q^4)$, in agreement with the
field theory Airy coefficients of~\cite{Bobev:2025ltz}. The one-loop
character also simplifies considerably. Setting $x=\mathrm e^{-Qt/3}$, so
that $q_1q_2=x^3$, gives
\begin{align}
 \mathcal K_{V^{5,2}}(t)
 &=\frac{x^3(1+x)}
 {(1-q_1)(1-q_2)(1-x)^4}
 \Bigl[
 q_1^{-1}-q_1+q_2^{-1}-q_2
 +5(x-x^{-1})+x^{-2}-x^2
 \Bigr].
 \label{eq:V52charSC}
\end{align}
Expanding $(1-x)^{-4}$ and Mellin transforming as in the ABJM case gives
a finite combination of ordinary six-fold Barnes zeta functions. Introducing the Barnes period
$\bm\Omega_6=(b,b^{-1},Q/3,Q/3,Q/3,Q/3)$ and
$\zeta_6(s,z)\equiv\zeta_6^{(0)}(s,z)$, we find
\begin{align}
 &\zeta_{V^{5,2}}(s)
 =\sum_{\epsilon=b,b^{-1}}\Bigl[
 \zeta_6(s,\epsilon)
 +\zeta_6\Bigl(s,\frac{Q}{3}+\epsilon\Bigr)
 -\zeta_6(s,Q+\epsilon)
 -\zeta_6\Bigl(s,\frac{4Q}{3}+\epsilon\Bigr)
 \Bigr]\nn\\
 &+\zeta_6\Bigl(s,\frac{Q}{3}\Bigr)
 -4\zeta_6\Bigl(s,\frac{2Q}{3}\Bigr)
 -5\zeta_6(s,Q)
 +5\zeta_6\Bigl(s,\frac{4Q}{3}\Bigr)
 +4\zeta_6\Bigl(s,\frac{5Q}{3}\Bigr)
 -\zeta_6(s,2Q)\, ,
 \label{eq:V52Barnes}
\end{align}
so that
\begin{equation}
 \cA_{V^{5,2}}(b,0,0)=\zeta'_{V^{5,2}}(0)\, .
 \label{eq:V52ASC}
\end{equation}
To our knowledge, no analytic expression for this real Airy constant was
previously available; the authors of~\cite{Bobev:2025ltz} instead determined it
numerically.

In order to compare to previous numerical results in the literature, we may look at the  small $b$ behaviour. Rescaling the equivariant parameter as $t=bu$ gives, with $x=\mathrm e^{-u/3}$,
\begin{equation}
 \lim_{b\to0}b^2\mskip2mu\mathcal K_{V^{5,2}}(bu)
 =\frac{1}{u}
 \frac{(1+x)^2(x^2+3x+1)}
 {(1-x)^2(1+x+x^2)}\, .
 \label{eq:V52smallbchar}
\end{equation}
Expanding the rational function in Dirichlet characters and Mellin
transforming gives precisely the result \eqref{eq:V52g0}. 

\section{Integrated correlators}
\label{app:correlate}

Derivatives of the sphere free energy
$F(b;m_1,m_2,m_3)\equiv-\log Z_{S^3_b}(m_1,m_2,m_3)$
with respect to $b$ and $m_i$ give integrated correlation functions.
In particular, derivatives evaluated at $b=1$ and $m_i=0$ give
integrated correlators of the undeformed ABJM SCFT.  Our generic Airy
result determines the perturbative large $N$ free energy for positive
integer $k$, arbitrary squashing, and all three independent real masses.

The $\mathcal N=6$ integrated-correlator analysis of
\cite{Binder:2019mpb} used the round-sphere two-mass sector,
parametrized by $m_\pm$, for which all-orders large $N$ Fermi gas
results were available.  It was noted that the third real mass gives
additional integrated constraints, 
but the analysis of \cite{Binder:2019mpb} was restricted to the
two-mass sector.
More generally, \cite{Chester:2021gdw} considered
$F(b;m_1,m_2,m_3)$ and showed that, at fourth order, one additional
independent mass derivative remains for generic $\mathcal N=6$ ABJM,
$\partial_{m_2}^2\partial_{m_3}^2F$, whose large $N$ expansion was
not known.

We parametrize real mass deformations around the ABJM superconformal point
in terms of the three real masses $m_i$ by
\begin{align}
\label{eq:ABJMmassmap}
    \Delta_1&= \frac{1}{2} - \ii \frac{m_1+m_2+m_3}{Q} \,, \qquad
    \Delta_2 = \frac{1}{2} -\ii\frac{m_1-m_2-m_3}{Q} \,, \nonumber\\
    \Delta_3 &= \frac{1}{2} + \ii \frac{m_1+m_2-m_3}{Q} \,, \qquad
    \Delta_4 = \frac{1}{2} + \ii \frac{m_1-m_2+m_3}{Q}\,,
\end{align}
as in \cite{Bobev:2025ltz}.  Following \cite{Chester:2021gdw}, we define
$m_\pm\equiv m_2\pm m_1$ and consider
\begin{align}
    \mathcal I_{23}
    &\equiv
    \left.\partial_{m_2}^2\partial_{m_3}^2F\right|_0 \,, \nn\\
    \mathcal J
    &\equiv
    \left.
    \left(
    2\partial_{m_+}^4F
    -2\partial_{m_+}^2\partial_{m_-}^2F
    \right)\right|_0 \,,
\end{align}
where $|_0$ denotes $b=1$, $m_i=0$.  The quantity $\mathcal J$ is
determined entirely within the two-mass slice $m_3=0$, which by
\eqref{eq:ABJMmassmap} is the Nosaka locus \eqref{eq:NosakaLocus},
whereas $\mathcal I_{23}$ is the additional independent fourth-order
mass derivative for generic $\mathcal N=6$ ABJM.  For $k=1,2$,
enhanced $\mathcal N=8$ supersymmetry implies
\cite{Binder:2018yvd,Chester:2021gdw}
$\mathcal I_{23}=\mathcal J$.

We now evaluate the complete perturbative large $N$ contribution to
$\mathcal I_{23}-\mathcal J$ at fixed positive integer $k$.
Remarkably, differentiating the complete Airy answer we find that the contributions from $\mathcal B$ and $\mathcal C$ cancel so that
\begin{align}\label{eq:F3}
\cD_k \equiv \left(\mathcal{I}_{23}-\mathcal{J}\right)^\text{pert}
&=
-\left.
\partial_{m_2}^2\partial_{m_3}^2
\mathcal A_{\rm ABJM}(k,1,\bm\Delta(m))
\right|_0
+k^4 A^{''''}(k)\,.
\end{align}
Thus the perturbative departure from the $\mathcal N=8$ relation at generic $k$ is
entirely an $N^0$ effect, corresponding to order $c_T^{-2}$ in the
normalized four-point function, namely the order at which one-loop bulk
effects first enter.

We can make the generic $k$ dependence of \eqref{eq:F3}
more explicit. 
Differentiating the Molien character before setting the masses to zero,
and combining the two terms in \eqref{eq:F3}, gives the simple
representation
\begin{align}
\cD_k
&=
\int_0^1 \frac{\diff x}{x}\,(-\log x)^3 H_k(x)\,,
\label{eq:DkIntegral}
\end{align}
where $x=\ex^{-t/2}$, with $t$ is the variable entering
the one-loop character, and 
\begin{align}
H_k(x)
&=
\frac{256x^4(1+x^2)}{k(1-x^2)}
\sum_{r=1}^{\lfloor(k-1)/2\rfloor}
\frac{\sin^2\theta_r}
{\bigl(1-2x\cos\theta_r+x^2\bigr)^4}\,,
\qquad
\theta_r=\frac{2\pi r}{k}\,,
\label{eq:HkRoots}
\end{align}
where
$\lfloor x\rfloor$ denotes the greatest integer less than or equal to
$x$. 
The sum is empty for $k=1,2$, while every summand is strictly
positive for $0<x<1$ and $k\geq3$. Hence
\begin{align}
\cD_1=\cD_2=0\,,
\qquad
\cD_k>0\quad\text{for every }k\geq3\,.
\end{align}
Thus, the enhanced $\mathcal N=8$ relation is manifest at
$k=1,2$, and its perturbative violation is strictly positive
at every higher level.
This representation also makes convergence transparent:
$H_k(x)=O(x^4)$ as $x\to0$, and $H_k(x)$ has at most a simple
pole as $x\to1$. The integral \eqref{eq:DkIntegral} therefore
converges without further regularization. Although we shall not use it here,
it is interesting to note that we can also write
\begin{align}
	H_k(x)
	&=
	\frac{32x^2(1+x^2)}{3}
	\Bigg[
	-\frac{k^3x^k(1+4x^k+x^{2k})}
	{(1-x^2)^3(1-x^k)^4}
	\nonumber\\
	&\hspace{21mm}
	+\frac{kx^k(1+10x^2+x^4)}
	{(1-x^2)^5(1-x^k)^2}
	+\frac{6x^2(1+x^2)(1+x^k)}
	{(1-x^2)^6(1-x^k)}
	\Bigg]\,.
	\label{eq:Hk}
\end{align}

The poles of $H_k(x)$ lie at roots of unity, and the integral
\eqref{eq:DkIntegral} can be evaluated analytically for every
positive integer $k$. With $\theta_r$ as defined above, set
\begin{align}
u_r&=1-\frac{2r}{k}\,,
&
s_r&=\sin\theta_r\,,
&
c_r&=\cos\theta_r\,,
\end{align}
for $r=1,\ldots,\lfloor(k-1)/2\rfloor$.
Then $\mathcal{D}_k$ can be evaluated in terms of residues at the poles of $H_k(x)$ which lie at $x=\pm 1$ and 
$x=\ex^{\pm\ii \theta_r}$, and we find
\begin{align}
\cD_k
&=
\frac{\pi^4}{15}B_+(k)
-\frac{7\pi^4}{120}B_-(k)
+\sum_{r=1}^{\lfloor (k-1)/2\rfloor}
\cF_r(k)\,,
\label{eq:DkExact}
\end{align}
where
\begin{align}
B_+(k)
&=
\frac{4(k^2-1)(k^2-4)(k^2+5)}
{945\,k}\,,
\\
B_-(k)
&=
\begin{cases}
B_+(k)\,, & k \ \text{even}\,,\\[4pt]
\dfrac{4k(k^2-1)(k^2+1)}{15}\,, & k \ \text{odd}\,,
\end{cases}
\end{align}
and
\begin{align}
\cF_r(k)
=
\frac{1}{k}\Bigg[
&16\pi u_r\,\frac{c_r}{s_r^3}
+24\pi^2
\left(u_r^2-\frac13\right)
\frac{1+2c_r^2}{s_r^4}-\frac{88\pi^3}{3}
u_r(1-u_r^2)
\frac{c_r(2+c_r^2)}{s_r^5}
\nonumber\\
&+\pi^4
\left(
4u_r^4-8u_r^2+\frac{28}{15}
\right)
\frac{1+6c_r^2+c_r^4}{s_r^6}
\Bigg]\,.
\label{eq:Fr}
\end{align}
The $B_\pm(k)$ terms arise from the poles at $x=\pm 1$, and $\cF_r(k)$ arise from the conjugate pairs $x=\ex^{\pm \ii \theta_r}$.
Thus \eqref{eq:DkExact} gives an exact finite expression for $\cD_k$ at positive integer Chern--Simons level $k$.
For example,
\begin{align}
\cD_3
&=
-\frac{64\sqrt{3}}{81}\pi
-\frac{128}{27}\pi^2
+\frac{2816\sqrt{3}}{729}\pi^3
-\frac{10504}{6561}\pi^4\,,
\nn\\
\cD_4
&=
\frac{\pi^4}{16}-\frac{\pi^2}{2}\,,
\end{align}
while numerically
\begin{align}
\cD_3&=0.4127447779563\ldots\,,
&
\cD_4&=1.1532659890805\ldots\,,
\nonumber\\
\cD_5&=2.1818979209498\ldots\,,
&
\cD_6&=3.4787118110751\ldots\,.
\end{align}

\bibliographystyle{utphys} 
\bibliography{references}{}

\end{document}